\documentclass[aps,prd,reprint,superscriptaddress,floatfix,nofootinbib,showkeys,showpacs,preprintnumbers]{revtex4-2}
\pdfoutput=1

\usepackage{tabularx}
\usepackage{xspace}
\usepackage{listings}
\usepackage{lineno}
\usepackage{bm}
\usepackage{amsmath}
\usepackage{amssymb}
\usepackage[dvipsnames]{xcolor}
\usepackage{graphicx}
\usepackage{comment}
\usepackage{natbib}
\usepackage{hyperref}
\usepackage{amsmath}
\usepackage{enumitem}
\setlist{wide, labelwidth=!, labelindent=0pt}
\usepackage{multirow}
\hypersetup{    
  colorlinks      = true,
  linkcolor       = {black},
  linkbordercolor = {white},
  citecolor       = {black},
  citebordercolor = {white},
  urlcolor        = {blue},
  urlbordercolor  = {white},
}

\usepackage{soul} 
\usepackage{amsmath}
\usepackage{amssymb}
\usepackage{xspace}
\usepackage{xifthen}
\usepackage{float}
\usepackage{mathtools}
\usepackage{relsize}
\usepackage{aas_macros}

\makeatletter 
\newcommand{\LCASES}[1]{$\m@th\displaystyle{#1}$\hfil}
\newcommand{\CCASES}[1]{\hfil$\m@th\displaystyle{#1}$\hfil}
\newcommand{\RCASES}[1]{\hfil$\m@th\displaystyle{#1}$}
\makeatother

\newcases{ecases}{\quad}{\CCASES{##}}{\LCASES{##}}{\lbrace}{.}
\newcases{ecases*}{\quad}{\CCASES{##}}{{##}\hfil}{\lbrace}{.}

\graphicspath{{Figures/}}

\def\beq{\begin{eqnarray}}
\def\eeq{\end{eqnarray}}

\def\siglogM{\sigma_{\log M}}
\def\Mmin{M_\mathrm{min}}

\def\Qcen{\mathcal{Q}_\mathrm{cen}}
\def\Qsat{\mathcal{Q}_\mathrm{sat}}

\def\Reff{R_\mathrm{eff}}

\begin{document}

\preprint{APS/123-QED}

\title[HOD-based forward modeling of voids]{Multiprobe cosmology forecasts from halo-occupation-distribution-based forward modeling of galaxy and void statistics}

\author{Andr\'{e}s N. Salcedo}
\affiliation{Steward Observatory, University of Arizona, 933 North Cherry Avenue, Tucson, AZ 85721, USA.}
\affiliation{Department of Physics, University of Arizona, 1118 East Fourth Street, Tucson, AZ 85721, USA.}
\email{ansalcedo@arizona.edu}
\author{Alice Pisani}
\affiliation{Aix-Marseille Universit\'{e}, CNRS/IN2P3, CPPM, Marseille, France.}
\affiliation{Department of Astrophysical Sciences, Peyton Hall, Princeton University, Princeton, NJ 08544, USA.}
\author{Nico Hamaus}
\affiliation{Universit\"ats-Sternwarte M\"unchen, Fakult\"at f\"ur Physik, Ludwig-Maximilians-Universit\"at, Scheinerstr. 1, 81679 M\"unchen, Germany}

\date{\today}

\begin{abstract}
The large under-dense regions in the cosmological matter density field, known as cosmic voids, are powerful probes of cosmology but their potential is currently under-exploited. Observationally, voids are identified within the large scale distribution of galaxies and are therefore sensitive to certain features of the galaxy-halo connection. This sensitivity makes the combination of void and galaxy summary statistics particularly powerful probes of both cosmology and the galaxy-halo connection through self-calibration of the void-galaxy relation. In particular the combination of void and galaxy summary statistics breaks degeneracies in the galaxy-halo connection and cosmology relative to the case of galaxy clustering alone. To demonstrate this we forecast cosmological constraints attainable from the combination of the void size function $n_v$, projected void-galaxy cross-correlation function $w_{p,vg}$ and projected galaxy auto-correlation function $w_{p,gg}$ measured in Dark Energy Spectroscopic Instrument (DESI) Year 5 data. We use a grid of cosmological N-body simulations to model this datavector as a function of $\sigma_8$, $\Omega_m$ and a flexible halo occupation distribution (HOD) model that includes central and satellite galaxy assembly bias. For our fiducial scenario combining $n_v$, $w_{p,vg}$ and $w_{p,gg}$ we forecast $1.5\%$ and $0.8\%$ constraints on $\Omega_m$ and $\sigma_8$ from DESI-Y5 data. We also forecast constraints from the combination of the void size function and stacked weak lensing signal of voids.
\end{abstract}

\maketitle

\section{Introduction}

Cosmic voids, the large under-dense regions of the cosmic web as traced by galaxies, are an underexploited but powerful probe of cosmology \citep[e.g.][]{Pisani_et_al_2019}. Within the standard $\Lambda\mathrm{CDM}$ paradigm void statistics provide independent and complementary constraints on the growth of structure (i.e. $\Omega_m$ and $\sigma_8$) to those from standard cosmological probes \citep[e.g.][]{Kreisch_et_al_2022, Contarini_et_al_2022, Pelliciari_et_al_2023}. Owing to their low-density interiors they also provide unique information on neutrino masses \citep[e.g.][]{Massara_et_al_2015, Banerjee_Dalal_2016, Kreisch_et_al_2019, Sahlen_2019, Schuster_et_al_2019, Contarini_et_al_2021, Kreisch_et_al_2022, Thiele_et_al_2024} as well as various extensions and deviations from standard $\Lambda\mathrm{CDM}$ such as dark energy theories \citep[e.g.][]{Spolyar_et_al_2013, Pisani_et_al_2015, Pollina_et_al_2016, Sahlen_et_al_2016, Falck_et_al_2018, Sahlen_Silk_2018, Paillas_et_al_2019, Perico_et_al_2019, Verza_et_al_2019, Contarini_et_al_2022, Verza_et_al_2023}, modified gravity \citep[e.g.][]{Clampitt_et_al_2013, Cai_et_al_2015, Hamaus_et_al_2015, Zivick_et_al_2015, Achitouv_2016, Sahlen_et_al_2016, Baker_et_al_2018, Contarini_et_al_2021}, and primordial non-Gaussianity \citep[e.g.][]{ChanKC_et_al_2019}.

One of the most extensively studied void statistics is the counts of cosmic voids as a function of their size, referred to as the void size function \citep[e.g.][]{Pisani_et_al_2015, Nadathur_et_al_2016, Sahlen_et_al_2016, Contarini_et_al_2019, Ronconi_et_al_2019, Verza_et_al_2019, Contarini_et_al_2021, Contarini_et_al_2022, Contarini_et_al_2023, Pelliciari_et_al_2023}. Two point functions of voids have also been studied in the literature, including the void-galaxy cross-correlation function \citep[][]{Lavaux_and_Wandelt_2012, Hamaus_et_al_2014d, Pisani_et_al_2014, Cai_et_al_2016, Hamaus_et_al_2016, Hamaus_et_al_2017, Hawken_et_al_2017, Achitouv_et_al_2019, Correa_et_al_2019, Hamaus_et_al_2020, Hawken_et_al_2020, Nadathur_et_al_2020, Aubert_et_al_2022, Hamaus_et_al_2022, Woodfinden_et_al_2022, Schuster_et_al_2023, Fraser_et_al_2024}, the void auto-correlation function \citep[e.g.][]{Hamaus_et_al_2014b, Hamaus_et_al_2014c, ChanKC_et_al_2014, Clampitt_et_al_2016, Chuang_et_al_2017, Kreisch_et_al_2019, Voivodic_et_al_2020, Kreisch_et_al_2022}, and void lensing \citep[e.g.][]{Krause_et_al_2013, Clampitt_et_al_2015, Chantavat_et_al_2016, Cai_et_al_2017, Chantavat_et_al_2017, Sanchez_et_al_2017, Baker_et_al_2018, Brouwer_et_al_2018, Fang_DES_voids_et_al_2019, Davies_et_al_2021, Vielzeuf_et_al_2021, Bonici_et_al_2023}. These statistics are typically modelled theoretically starting from expectations for voids in the matter distribution which must then be linked to voids in the tracer (i.e. galaxy) distribution. In the case of the void size function the standard modeling approach \citep[e.g.][]{Sheth_vdWeygaert_2004, Jennings_et_al_2013} is based on the excursion-set theory \citep{ps:74} with an additional empirical prescription for the void-tracer relation that is often calibrated using simulations \citep[e.g.][]{Contarini_et_al_2019,Contarini_et_al_2021,Contarini_et_al_2022, Contarini_et_al_2023, Contarini_et_al_2024}. Similarly the standard approach to modeling the void-galaxy cross-correlation function \citep[e.g.][]{Hamaus_et_al_2014, Schuster_et_al_2023} assumes linear theory and that stacked voids are spherically symmetric \citep{Ryden_Melott_1996, Lavaux_and_Wandelt_2012} with an additional linear bias relation between matter and galaxy voids. 

Within any theoretical framework for modeling cosmic voids two related challenges present themselves. The first is that theory can straightforwardly predict the statistics of voids in the matter distribution, but since voids are observationally identified in the galaxy distribution some model of the void-tracer relation must be included and either calibrated or marginalized over. The second challenge is that voids are observationally identified with a specific void finder that in principle produces its own particular selection effects and complicates the connection between ``true'' matter voids and voids found in the galaxy distribution.

Simulation-based forward modeling can address both of these challenges by modeling the galaxy-halo connection of the galaxies used to observationally identify voids and then consistently applying the same void-finder used in the data to the simulated galaxy population to predict relevant void statistics. In this paper we chose to focus on the halo occupation distribution \citep[e.g.][]{Berlind_2002, Zheng_et_al_2005} framework for modeling the galaxy-halo connection, though we note alternative frameworks such as subhalo abundance matching \citep[SHAM; e.g.][]{vo:04, Conroy_et_al_2006} and its extensions \citep[e.g.][]{Contreras_et_al_SHAMe_2021, Ortega-Martinez_et_al_2024} as well as semi-analytic models could similarly be applied to voids. The HOD framework specifies the statistical connection between the properties of halos and their galaxy content. Within a simulation-based forward-modeling framework an HOD model can in principle predict the void population identified from a given galaxy sample at no additional cost in terms of model parameters. In addition to modeling all void summary statistics within a single framework this allows for joint modeling of void and galaxy statistics in order to self-calibrate the void-galaxy relation.

In this paper we investigate the cosmological constraints obtainable by applying this approach to voids identified in spectroscopic galaxy data. Specifically we investigate the constraints on $\sigma_8$ and $\Omega_m$ that can be obtained by analyzing the void size function $n_v$, the projected void-galaxy cross-correlation function $w_{p,vg}$, the void excess surface density profile $\Delta \Sigma_{vm}$ as measured by weak lensing, and the projected galaxy auto-correlation function $w_{p,gg}$. We compute Fisher forecasts for various combinations of these observables assuming a spectroscopic survey with properties similar to the final data release of the Dark Energy Spectroscopic Instrument survey. To compute the signal for our multi-probe datavector we populate N-body halos from the AbacusSummit suite of cosmological simulations using an extended HOD formalism and identify voids in the HOD galaxy distribution using the $V^2$ void finding algorithm.

The next section describes in detail our methodology for creating mock void and galaxy catalogs and measuring our datavector from them. In Section \ref{sec:methods} we describe our numerical simulation suite, HOD methodology, void-finding methodology, and we define our multi-probe summary statistics. In Section \ref{sec:sens} we investigate the sensitivity of our mock datavectors to HOD and cosmological parameters. Section \ref{sec:forecasts} describes how we compute Fisher forecasts with our mock datavectors, including how we estimate relevant covariance matrices, and also presents our main results. Finally in Section \ref{sec:conc} we summarize our findings and identify directions for future work.

\section{Galaxy and Void Modeling}
\label{sec:methods}

\subsection{Simulations and halo identification}
\label{subsec:sims}

We use halo and particle data from the AbacusSummit\footnote{{https://abacussummit.readthedocs.io/en/latest/index.html}} \citep{Maksimova_Summit_et_al_2021} suite of N-body simulations in our analysis. The AbacusSummit simulations are run with the {\sc{abacus}} \citep{Metchnik_2009, Garrison_et_al_2018, Garrison_et_al_2019, Garrison_et_al_2021} cosmological N-body code which uses GPUs and novel computational techniques to achieve high speed and accuracy. Specifically we use 25 realizations of the fiducial AbacusSummit cosmology, based on \citet{Planck_2016} results as well as the $\Omega_m$ and $\sigma_8$ variations from the derivative grid (see Table \ref{tab:cosmo}). All simulations used are periodic cubes with side-length $L_\mathrm{side} = 2.0 \, h^{-1} \, \mathrm{Gpc}$, $N_\mathrm{part} = 6192^3$ particles of mass $M_\mathrm{part} \sim 2 \times 10^9 \, h^{-1} \, M_\odot$, and spline force softening length $\epsilon_g = 7.2 \, h^{-1} \, \mathrm{kpc}$ (as described in \citet{Garrison_et_al_2018}). In what follows we utilize simulation redshift snapshots at $z = 0.5$.
 
Haloes are identified from particle snapshots using the {\sc{CompaSO}} halo finder \citep{Hadzhiyska_COMPASO_et_al_2022}. The {\sc{CompaSO}} algorithm extends existing spherical overdensity (SO) algorithms by considering the tidal radius around small haloes in the vicinity of larger neighbours to competitively assign particles. By doing this {\sc{CompaSO}} more effectively deblends nearby haloes. {\sc{CompaSO}} roughly proceeds in three steps, Level 0 (L0) haloes are first identified with a modified friends-of-friends algorithm, then Level 1 (L1) haloes are identified with a competitive assignment SO algorithm, and finally Level 2 (L2) haloes are idenfitied in a similar way but with a higher density threshold. In this schema L0 groups are large overdensities of matter that may contain multiple L1 haloes which correspond to distinct haloes in a standard SO-algorithm, and L2 haloes correspond to subhaloes. For our purposes we use `cleaned' {\sc{CompaSO}} halo catalogues and adopt as the halo center for each of our L1 haloes the center-of-mass of their most massive embedded L2 halo, as recommended in \citet{Hadzhiyska_COMPASO_et_al_2022}.

\begin{table}
   \centering
   \caption{Cosmology and simulation parameters. Simulations are listed as named in AbacusSummit documentation where each $\mathrm{cXXX}$ refers to a different cosmology.}
    \begin{tabular}{cccccc}      
    \hline
    Parameter & c000 & c102 & c103 & c112 & c113 \\
    \hline
    $\Omega_m$ & $0.314$ & $0.336$ & $0.293$ & $0.314$ & - \\
    $\Omega_\Lambda$ & $0.686$ & $0.664$ & $0.707$ & $0.686$ & - \\
    $h$ & $0.674$ & $0.660$ & $0.688$ & $0.674$ & - \\
    $\sigma_8$ & $0.808$ & - & - & $0.824$ & $0.792$ \\
    $n_s$ & $0.965$ & - & - & - & - \\
    $w$ & $-1.0$ & - & - & - & - \\
    \hline
   \end{tabular}
\label{tab:cosmo}
\end{table}

\subsection{Halo occupation distribution modeling of void tracer galaxies}
\label{subsec:HOD}

To model galaxy identified voids we populate our simulated haloes with galaxies according to a halo occupation distribution (HOD) framework \citep[e.g.][]{Jing_Mo_Borner_1998, Benson_et_al_2000, Ma_Fry_2000,Peacock_Smith_2000, Seljak_2000, Scoccimarro_et_al_2001, Berlind_2002, Cooray_Sheth_2002, Yang_et_al_2003, vdBosch_et_al_2003b, Zheng_et_al_2005, Cooray_2006, Mandelbaum_et_al_2006, Zheng_2009, Zehavi_et_al_2011, Coupon_et_al_2012, Leauthaud_et_al_2012, Guo_2014, Zu_Mandelbaum_12-2015, Zehavi_et_al_2018}. We parametrize the mean central and satellite occupations of our haloes with the widely used equations of \citet{Zheng_et_al_2005},
\begin{align}
\langle N_\mathrm{cen} | M_h \rangle &= \frac{1}{2} \left[1 + \mathrm{erf}\left( \frac{\log M_h - \log \Mmin}{\siglogM} \right) \right], \label{eq:cen_hod} \\
\langle N_\mathrm{sat} | M_h \rangle &= \langle N_\mathrm{cen} | M_h \rangle \left( \frac{M_h - M_0}{M_1} \right)^\alpha,
\end{align}
where the parameter $\Mmin$ represents the characteristic minimum mass of halos that host a central ($\langle N_\mathrm{cen} | \Mmin \rangle = 0.5$) and the parameter $\siglogM$ represents the width of the transition from $\langle N_\mathrm{cen} \rangle = 0$ to $\langle N_\mathrm{cen} \rangle = 1$. The parameter $M_1$ is the characteristic mass for a halo to host its first satellite, $\alpha$ is the slope of the satellite-occupation power law, and $M_0$ is a cutoff to this power law.

The actual number of centrals and satellites placed into each halo is drawn randomly from a binomial and Poisson distribution respectively with mean occupations given above. Centrals are placed at the center of their host halo, while satellites are distributed according to a Navarro-Frenk-White profile \citep[NFW;][]{NFW_1997} parameterized by halo concentration $c_h = r_h / r_s$. We use the fits of \citet{Correa_2015} to assign concentrations to our simulated haloes. 

Following \citet{Salcedo_et_al_2022b} \citep[see also][]{McEwen_2018, Wibking_et_al_2019, Salcedo_et_al_2022, Xu_et_al_2021} we extend our HOD to allow for galaxy assembly bias. Galaxy assembly bias refers to the possibility for galaxy occupation at fixed host halo mass to depend on secondary halo and environmental properties. Because voids probe the lowest density regions of the cosmic web they may be particularly sensitive to galaxy assembly bias. This presents both the challenge of robustly marginalizing over possible assembly bias effects, but also the opportunity to exploit new valuable information on galaxy formation. We allow the central and satellite occupations to vary on a halo-by-halo basis based on the matter overdensity measured in a top-hat sphere of radius $8.0 \, h^{-1} \, \mathrm{Mpc}$ centered on each individual halo $\delta_8^m$. This environmental dependence is written as,
\begin{align}
\log \Mmin &= \log M_\mathrm{min,0} + \Qcen ( \tilde{\delta}_8^m
 - 0.5 ) \\
\log M_1 &= \log M_{1,0} + \Qsat ( \tilde{\delta}_8^m
 - 0.5 )
 \end{align}
where $\Qcen$ and $\Qsat$ express the strength of the dependence of $\Mmin$ and $M_1$ respectively on the large scale environment and $\tilde{\delta}_8^m \in [0,1]$ is the normalized rank of $\delta_8^m$ within a narrow mass bin. 

For our fiducial HOD we choose parameter values of $\siglogM = 0.6$, $\log \Mmin = 13.3$, $\log M_0 = 11.0$, $\log M_1 = 14.2$, $\alpha = 1.50$, $\Qcen = 0.0$, and $\Qsat = 0.0$. The resulting galaxy number density is $n_g \sim 3 \times 10^{-4} \, h^3 \, \mathrm{Mpc}^{-3}$ with mean galaxy separation of $\sim 15.0 \, h^{-1} \, \mathrm{Mpc}$. These parameters were chosen to be consistent with a Luminous Red Galaxy (LRG) sample identified in either SDSS or DESI data.

\subsection{Void finding in HOD galaxy catalogs}
\label{subsec:void_find}

To identify voids in our mock galaxy catalogs we use the {$\mathrm{V^2}$} software package which is based off of the {\sc{zobov}} algorithm \citep{Neyrinck_ZOBOV_2008}. This implementation is included within the publically available Void Analysis Software Toolkit (VAST)\footnote{https://github.com/DESI-UR/VAST}. The {\sc{zobov}} algorithm first creates a Voronoi tessellation of the galaxy population and uses a watershed algorithm to construct voids out of these cells \citep{Neyrinck_ZOBOV_2008}. Zones are constructed around each density minimum using a watershed algorithm. Density minima are Voronoi cells with lower density than all of their neighbours. Each Voronoi cell is grouped with its lowest density neighbour. Zones are then joined using a watershed algorithm according to some convention. 

We adopt the original convention of the VIDE void finder \citep{Sutter_et_al_2012} whereby adjacent zones are only added to a void if the density of the boundary between them is less than 0.2 times the mean tracer (galaxy) density. We choose to allow merging in order to increase our sensitivity to the galaxy-halo connection within intra-void filaments. This illustrates the power of a forward-modeling approach to void cosmology, within this framework it is straightforward to model the void statistics of any void finder and therefore we have complete freedom to make choices to enhance our sensitivity to our model parameters. The effective radius of the resulting voids is defined as,
\beq
\Reff = \left( \frac{3}{4 \pi} V \right)^{1/3},
\eeq
where $V$ is the total volume of the void's constituent Voronoi cells.

In what follows we analyze the statistics of voids found using the real-space positions of our HOD galaxies and therefore omit the impact of galaxy peculiar velocities and the Hubble flow. Both effects will displace the positions of tracer galaxies \citep[e.g.][]{Lavaux_and_Wandelt_2012, Hamaus_et_al_2020} and in principle provide additional cosmological information that our forecasts will leave out. It is plausible that the inclusion of these effects will improve our constraints, especially on $\Omega_m$, as AP effects add information on the expansion history while redshift-space distortions help to constrain the growth of structure, though these gains in information may be balanced by the increased nuisance parameter space needed to model our data vector in redshift-space. 

Galaxy peculiar velocities are relatively challenging to model at small scales. Central galaxies are not necessarily at rest with respect to their halo centers and satellite galaxy velocities can differ from those of the halo host's dark matter. In the HOD framework this behavior is typically modeled by including prescriptions for both central and satellite galaxy velocity bias \citep[e.g.][]{Guo_et_al_2015}. Galaxy velocity bias can in principle have some impact on the identification of voids in redshift space, but we expect that this impact will be minimal because its impact is most significant on small scales compared to the mean separation of our LRG-like galaxies of ${\sim}15\,h^{-1}\,\mathrm{Mpc}$, and because the change in galaxy velocities is not coherent.

\subsection{Summary statistics}
\label{ref:sum-stats}

In what follows we examine the void size function $n_v (R_v)$\footnote{Previous studies typically consider $ \frac{d n_v}{d \ln R_v}$.} (comoving space density of voids as a function of their radius $R_v$), projected void-galaxy cross-correlation function $w_{p,vg}$, void-galaxy lensing $\Delta \Sigma_{vm}$, and projected galaxy auto-correlation function $w_{p,gg}$. We compute the void size function $n_v$ in 9 equal linearly spaced bins $10.0 \, h^{-1} \, \mathrm{Mpc} < \Reff < 100.0 \, h^{-1} \mathrm{Mpc}$.

We use {\sc{corrfunc}} \citep{Sinha_2017} to compute the real-space void-galaxy cross-correlation function $\xi_{vg}(r_p, \pi)$ in 20 equal logarithmically spaced bins of $r_p$ covering scales $15.0 < r_p < 150.0 \, h^{-1} \, \mathrm{Mpc}$ and 200 equal linearly spaced bins out to $\Pi_\mathrm{max} = 200.0 \, h^{-1} \, \mathrm{Mpc}$.\footnote{We consider this value to be conservative as it is large relative to typical small-scale redshift space distortions (RSD). Though our analysis does not include RSD but we opt to be consistent with the choices one would make when analyzing real data.} This real-space correlation function is used to calculate the more observationally motivated quantity $w_{p,vg}$,
\beq
w_{p,vg}(r_p) = 2 \int^{\Pi_\mathrm{max}}_{0} d \pi \, \xi_{vg}(r_p, \pi),
\eeq
where $\Pi_\mathrm{max}$ refers to the maximum line-of-sight integration length.

To compute the void-lensing signal we again use {\sc{corrfunc}} to compute the real-space void-matter cross-correlation function $\xi_{vm}(r_p, \pi)$ in 60 equal logarithmically spaced bins of $r_p$ covering scales $0.2 < r_p < 200.0 \, h^{-1} \, \mathrm{Mpc}$ and 200 equal linearly spaced bins out to $\Pi_\mathrm{max} = 200.0 \, h^{-1} \, \mathrm{Mpc}$ which is then converted to the excess surface density,
\beq
\Delta \Sigma_{vm} (r_p) = \rho_m \left[ \frac{2}{r_p^2} \int_0^{r_p} r' w_{p,vm}(r') dr' - w_{p,vm} (r_p) \right],
\eeq
in 25 equal logarithmically spaced bins of $r_p$ covering scales $2.0 < r_p < 200.0 \, h^{-1} \, \mathrm{Mpc}$, where $\rho_m = \Omega_m \rho_\mathrm{crit}$ is the mean matter density. The projected galaxy clustering $w_{p,gg}$ is computed in 20 equal logarithmically spaced bins of $r_p$ covering scales $0.3 < r_p < 30.0 \, h^{-1} \, \mathrm{Mpc}$. 

\section{Sensitivity to HOD and cosmological parameters}
\label{sec:sens}

We examine the sensitivity of the void size function $n_v$, void-galaxy lensing $\Delta \Sigma$, and projected void-galaxy cross-correlation function $w_{p,vg}$ to our HOD and cosmological parameters. In the case of the two-point statistics $\Delta \Sigma$ and $w_{p,vg}$ we examine the sensitivity in two bins of void size $\Reff \in [20, 60)$ (small) and $\Reff \in [60, 100)$ (large). These bins have void number density of ${\sim}1.6 \times 10^{-6} \, h^{3} \, \mathrm{Mpc}^{-3}$ and ${\sim}1.3 \times 10^{-7} \, h^{3} \, \mathrm{Mpc}^{-3}$ respectively. For a similar sensitivity analysis of projected galaxy clustering $w_{p,gg}$ see \citet{Salcedo_et_al_2020}. We also investigate the sensitivity of the projected void autocorrelation function in Appendix \ref{app:wpvv}. 

\subsection{Sensitivity analysis of the void size function}
\label{subsec:sens_vsf}

\begin{figure*}
\centering \includegraphics[width=1.0\textwidth]{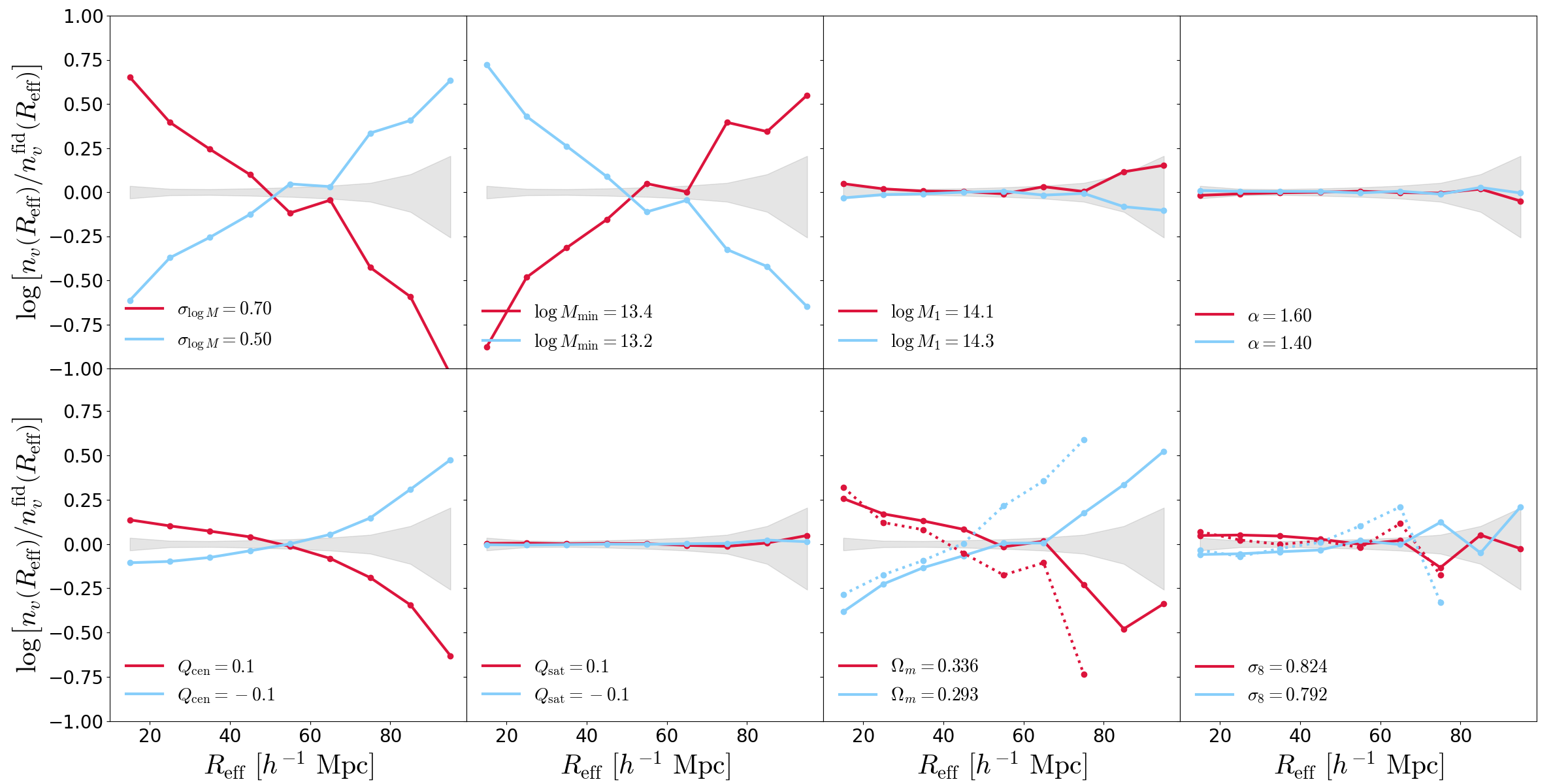}
    \caption{Sensitivity of void size function to our HOD and cosmological parameters. Each panel shows the logarithmic ratio of variations in each of our parameters against our fiducial model. Grey bands show uncertainties scaled to expectations from DESI-Y5 data (see Section \ref{subsec:cov}). In the panels for $\Omega_m$ and $\sigma_8$ dashed lines show the impact of parameter shifts on the matter voids.} 
\label{fig:VSF_vary}
\end{figure*}

We begin by considering the effects of parameter changes on $n_v$. Figure \ref{fig:VSF_vary} shows the impact of fixed variations in parameters for $n_v$, in each panel red (blue) curves show the effect of increasing (decreasing) the indicated parameter relative to our fiducial value. Grey bands represent the uncertainty expected from DESI-Y5 observations (described in detail in Section \ref{subsec:cov}). 

Beginning with $\siglogM$ we see that increasing (decreasing) $\siglogM$ increases (decreases) the abundance of voids with $\Reff \lesssim 50 \, h^{-1} \, \mathrm{Mpc}$ relative to the fiducial parametrization, while decreasing (increasing) the abundance of voids $\Reff \gtrsim 50 \, h^{-1} \, \mathrm{Mpc}$. This is because increasing $\siglogM$ boosts the occupation of haloes with $\log M < \log \Mmin$ that are more likely to be in underdense regions. The central galaxies these low mass haloes receive when increasing $\siglogM$ act to fracture voids that would otherwise have large $\Reff$ into multiple smaller voids. Turning to $\Mmin$ we observe similar behavior, decreasing (increasing) $\Mmin$ increases (decreases) the abundance of small voids and decreases (increases) the abundance of large voids. The parameter $\Mmin$ sets the characteristic mass to host a central galaxy, therefore decreasing it increases the number of low mass haloes that host a galaxy and fractures large voids into multiple smaller voids. This behavior is captured by topological void finders, which are therefore particularly sensitive to details of the galaxy-halo connection.

We now turn to $M_1$ and $\alpha$ which control the occupation of satellite galaxies. We see that increasing (decreasing) $M_1$ slightly increases the abundance of small and large voids and has very little impact on voids in the range $\Reff \sim 40 {-} 60 \, h^{-1} \, \mathrm{Mpc}$. By contrast variations in $\alpha$ have practically no effect on the void size function. This is unsurprising, $\alpha$ is the power-law index of the satellite occupation and therefore most strongly impacts the occupation of haloes with $M > M_1$. Regardless of their satellite occupation these massive haloes will host central galaxies and are highly clustered with other massive haloes that also host centrals. These haloes will never reside inside a potential void region and so their satellite occupation will have very little impact on the abundances of voids.

The parameter $\Qcen$ controls the amount of central galaxy assembly bias. We observe that when $\Qcen$ decreases (increases) the number of small voids is decreased (increased), while the number of large voids is increased (decreased). When $\Qcen$ is negative halos regions that are underdense for their mass are {\it less} likely to host central galaxies that would otherwise fracture large voids into multiple small voids. This suppresses the void size function at small $\Reff$ and enhances it for large $\Reff$. Similarly when $\Qcen$ is positive halos in underdense regions for their mass are more likely to host centrals. Relative to the case of $\Qcen = 0.0$ these centrals in underdense regions have the effect of breaking large voids into multiple smaller ones. The parameter $\Qsat$ controls the amount of satellite galaxy assembly bias and therefore has little impact of the void size function.

Finally we examine the sensitivity of the void size function to changes in the cosmological parameters $\Omega_m$ and $\sigma_8$. In these panels we also include the analogous sensitivity of matter voids in dashed lines. In all cases we downsample the particles so that they have the same space density as the fiducial galaxies in the fiducial cosmology. Beginning with $\Omega_m$ we see that an increase in the parameter enhances the abundance of small voids and suppresses that of large voids. This is because, at fixed $\sigma_8$, increasing $\Omega_m$ shifts power in the linear matter power spectrum from large to small scales relative to the normalization at $8\,h^{-1}\,\mathrm{Mpc}$ \citep{Pelliciari_et_al_2023, Contarini_et_al_2023}. Increasing $\Omega_m$ has a similar impact on the halo mass function but halos in the mass range that is suppressed are extremely rare and therefore don't contribute significantly to cosmological constraining power.  In contrast voids span a relatively large range in sizes and are therefore able to probe this behavior. 

This effect applies to matter voids which are in turn traced by galaxies. Shifting $\Omega_m$ at fixed HOD parameters {\it also} impacts galaxy occupation statistics. For example when $\Omega_m$ is increased at fixed $M_\mathrm{min}$ this is effectively the same as a decrease in $\Mmin$ at fixed $\Omega_m$ because all halo masses are increased on average (both variations increase the difference in the numerator of equation \ref{eq:cen_hod}). This effect contributes to the purely cosmological impact of shifting $\Omega_m$.  Turning to $\sigma_8$ we see that an increase in the parameter increases the abundance of small voids and has little effect on large voids. This makes sense as an increase in $\sigma_8$ at fixed $\Omega_m$ increases the matter power spectrum at all scales and therefore enhances the growth of matter perturbations. 

We emphasize that this sensitivity analysis is for voids identified in real-space. We do not expect our qualitative results to change in redshift space, though quantitative chances in sensitivity are likely to change. In particular we note the inclusion of realistic Alcock-Paczy\'nski \cite[AP; ][]{Alcock_Paczynski_1979} effects may actually lead to improvements in cosmological sensitivity, particularly on $\Omega_m$.

\subsection{Sensitivity analysis of void two-point functions}
\label{subsec:sens_2pt}

\begin{figure*}
\centering \includegraphics[width=1.0\textwidth]{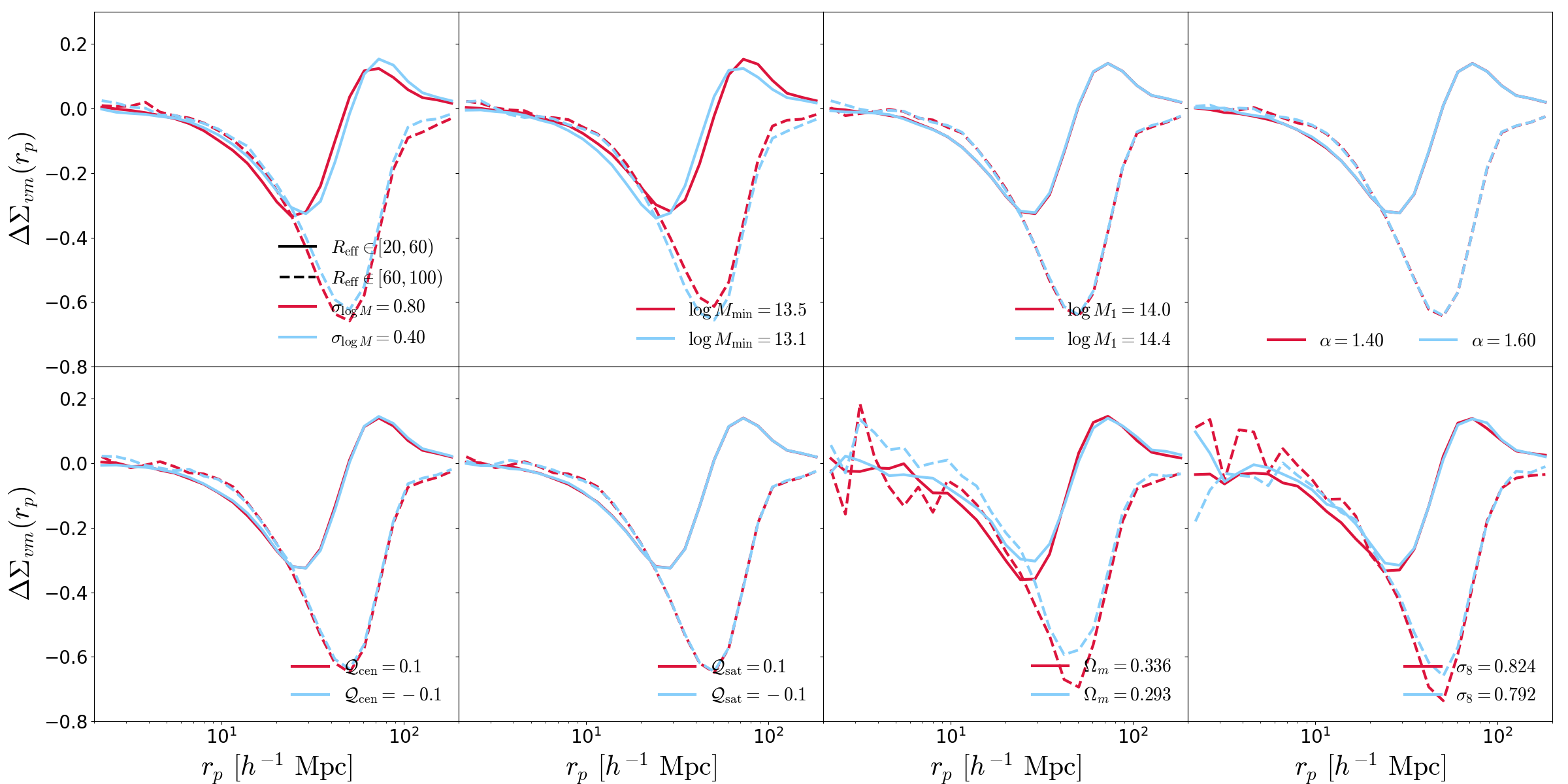}
    \caption{Sensitivity of void lensing to our HOD and cosmological parameters in two bins of void size, $R_v \in [ 20, 60 \, h^{-1} \, \mathrm{Mpc})$ (solid) and $R_v \in [ 60, 100 \, h^{-1} \, \mathrm{Mpc}]$ (dashed). Each panel shows variations in an HOD or cosmological parameter around our fiducial model. We do not plot uncertainties bands in this case since they will depend on the overlapping imaging to compute lensing, see Section \ref{sec:lens} for a description of our lensing covariances and see Figure \ref{fig:shape} for comparisons of our mean signal to either the sample variance or shape noise contributions to the lensing covarianace.} 
\label{fig:DSvm_vary}
\end{figure*}

Figure \ref{fig:DSvm_vary} shows the effect of changes in HOD and cosmological parameters on the void lensing profile of small ($\Reff \in [20,60)$, solid lines) and large ($\Reff \in [60, 100)$, dashed lines). We do not plot observational uncertainties here because they depend on what overlapping imaging is used to compute lensing observables, Figure \ref{fig:shape} shows the comparison of our fiducial signal with lensing uncertainties for a variety of scenarios which we forecast.

Beginning with $\siglogM$ ($\Mmin$) we see that an increase (decrease) in the parameter affects the small voids by slightly suppressing the overall profile and shifting the location of the maxima and minima to smaller radii. For the large voids there is a slight suppression of the overall profile, with a very negligible shift in extrema. In Figure \ref{fig:VSF_vary} we observed that the same changes in $\siglogM$ and $\Mmin$ increased the abundance of small voids and decreased that of large voids. Because the void size function is decreasing, a change in HOD parameters that fractures larger voids into smaller voids will fracture more small voids than large voids. Because these small voids inhabit lower bias environments than their large counterparts this fracturing will up-sample these lower bias environments and suppress the overall void bias. 

This fracturing is due to galaxies beginning to populate halos within intra-void filaments and sheets, and therefore the shifts observed in the lensing profile are due to the upweighting of these intra-void boundaries in the void-matter correlation function. This behavior depends on our choice to allow merging of local minima zones during void finding. By setting a density criterion for the merging of these zones we introduce sensitivity to the galaxy-occupation in these intra-void filaments.

We next turn to $M_1$, $\alpha$, $\Qcen$, and $\Qsat$. In the case of each parameter we observe negligible sensitivity of the void lensing profile. In the case of $\alpha$, $\Qcen$ and $\Qsat$ this is unsurprising as these parameters have negligible impact on the void size function, and therefore do not change the void populations in our two bins. On the other hand the behavior with respect to $\Qcen$ is quite perplexing as the parameter has a significant impact on the void size function. When $\Qcen$ is increased centrals are more (less) likely to be hosted in underdense (overdense) environments and the number of small voids increases similar to the cases described above. Therefore it is surprising but possible that the resulting suppression in the void bias is cancelled out by the tendency for voids to be located in relatively higher density environments.

Finally we consider the sensitivity of void lensing to our cosmological parameters. In both cases we see that parameter shifts have little impact on the location of the profile extrema but rather change the value of the minima. We see an increase in $\Omega_m$ increases our void lensing signal as expected since this increases the growth rate and increases the rate of matter evacuation from the voids. Likewise we also see that an increase in $\sigma_8$ increases our void signal since such an increase enhances all the features of the cosmic web. In the case of $\Omega_m$ we don't observe a discernible difference between the sensitivity of the small and large void samples, while the large voids appear to be more sensitive to $\sigma_8$ than their small counterparts. 

\begin{figure*}
\centering \includegraphics[width=1.0\textwidth]{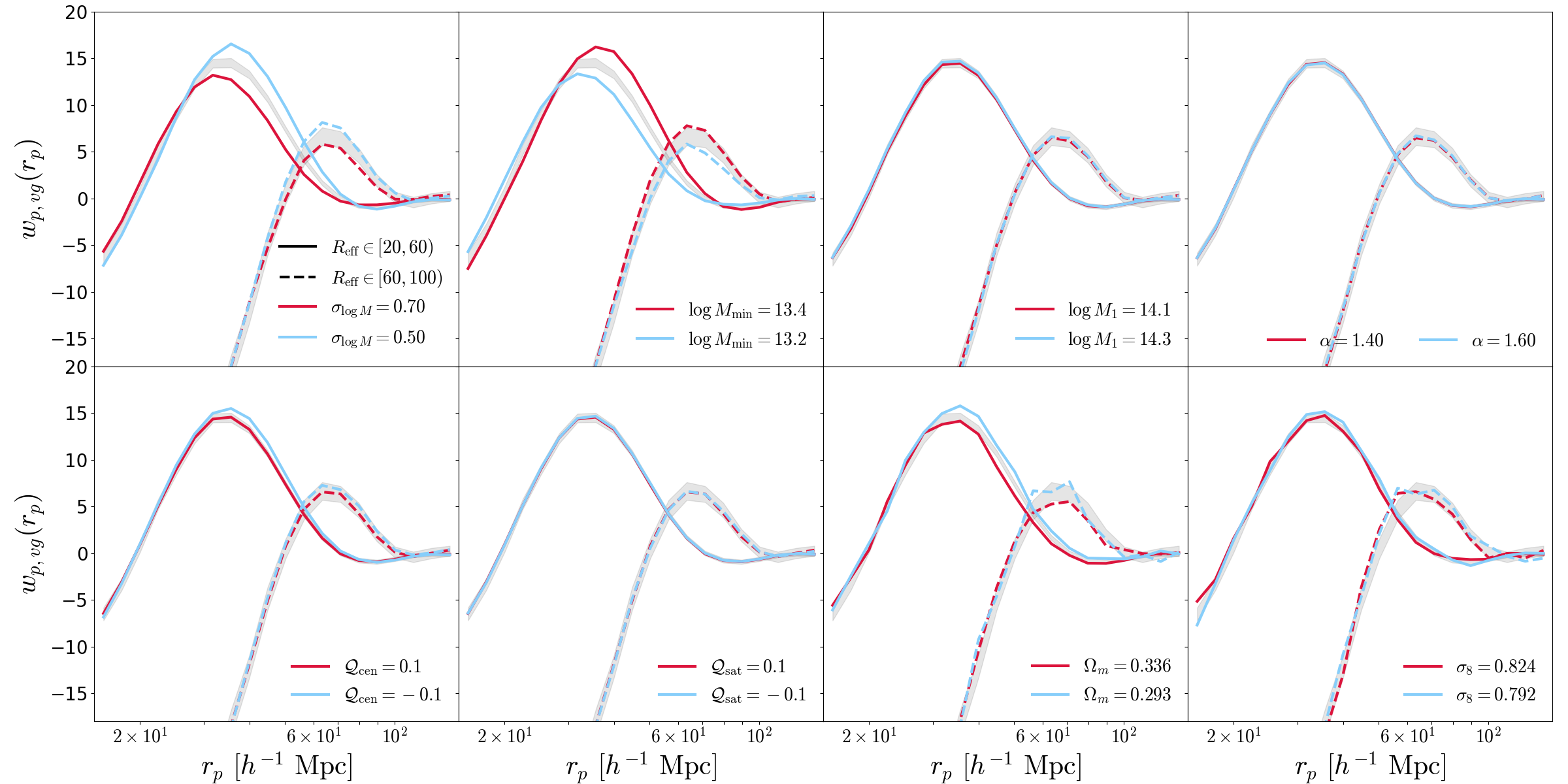}
    \caption{Sensitivity of the projected void-galaxy cross-correlation function to our HOD and cosmological parameters in two bins of void size, $R_v \in [ 20, 60 \, h^{-1} \, \mathrm{Mpc})$ (solid) and $R_v \in [ 60, 100 \, h^{-1} \, \mathrm{Mpc}]$ (dashed). Each panel shows variations in an HOD or cosmological parameter around our fiducial model. Grey bands show uncertainties scaled to expectations from DESI-Y5 data (see Section \ref{subsec:cov}). Note that the y-axis range has been restricted to highlight sensitivity of $w_{p,vg}$ at the separation of the void boundary, there is negligible sensitivity to either HOD or cosmology parameters in the suppressed range.} 
\label{fig:wpvg_vary}
\end{figure*}

In Figure \ref{fig:wpvg_vary} we show the effect of changes in HOD and cosmological parameters on the projected void-galaxy cross-correlation function of small ($\Reff \in [20,60)$, solid lines) and large ($\Reff \in [60, 100)$, dashed lines) with grey bands corresponding to uncertainties expected from DESI-Y5 observations. The behavior that we observe is similar to that observed in Figure \ref{fig:DSvm_vary} with some differences. Beginning with $\siglogM$ ($\log \Mmin$) we see that an decrease (increase) in the parameter significantly increases the amplitude of void-galaxy clustering for both the small and large voids. This can be explained by the resulting change in the galaxy bias $b_g$, decreasing (increasing) the parameter $\siglogM$ ($\Mmin$) has the effect of increasing the mean mass and bias of halos that host galaxies. For the small voids we also see a shift in the peak of the void-galaxy clustering that is consistent with the observed change in the void size function. Interestingly we do not observe a significant change in peak location in the case of the large void sample. This is consistent with what we observe in the analogous panel of Figure \ref{fig:VSF_vary} where variations in $\siglogM$ ($\Mmin$) have minimal impact from $60 {-} 70\,h^{-1}\,\mathrm{Mpc}$. Since the majority of the voids in this large size bin are in this size range this may account for the peak location remaining relatively fixed.

Turning to the parameters that control the satellite occupation: $M_1$, $\alpha$, and $\Qsat$ we observe negligible sensitivity in the void-galaxy clustering. Naively this may seem surprising since each of these parameters can modify $b_g$. However void-galaxy clustering also depends on the cross-correlation coefficient between voids and galaxies $r_{vg}$. Since satellites are primarily located in the highest density nodes of the cosmic web we interpret this lack of sensitivity as arising from a cancellation of changes in $b_g$ and $r_{vg}$. The impact of our central assembly bias parameter $\Qcen$ is primarily a change in the amplitude of void-galaxy clustering coming from a change in $b_g$. We observe negligible shifts in the location of the profile extrema when changing $\Qcen$.

We observe a decrease in $\Omega_m$ corresponds to an increase in void-galaxy clustering. This is due to the resulting increase in the galaxy bias. This increase is due to both a shift in halo masses relative to fixed values of $\Mmin$ and $M_1$ as well as a suppression of power in the large matter correlation function with increasing $\Omega_m$. In the case of $\sigma_8$ we observe minimal sensitivity of void-galaxy clustering. We again note that parameter sensitivity will in principle change in redshift-space, and may even be stronger given the inclusion of the AP effect.

\section{Cosmological Forecasts}
\label{sec:forecasts}

\subsection{Fisher Information and Forecasting}
\label{subsec:fisher}

Following the standard approach\footnote{Assuming a maximum likelihood estimator and Taylor series expanding the log-likelihood to second order.} to Fisher matrix forecasting \citep[e.g.][]{Tegmark_1997, DodelsonModernCosmology_2003, DarkEnergyTaskForce_Albrecht_2009} we write the Fisher information matrix as,
\beq
F_{ij} = \sum_{m,n} \frac{\partial \mathcal{O}(r_n)}{\partial \theta_i} C^{-1} (r_m, r_n) \frac{\partial \mathcal{O}(r_m)}{\partial \theta_j} ,
\eeq
where $C^{-1}$ refers to the inverse covariance matrix for observable $\mathcal{O}$ and $\theta$ is a vector of model parameters, in this case HOD and cosmology parameters. Partial derivatives of our observables with respect to our parameters are calculated using finite differences centered on our fiducial model. We smooth the measured derivatives for each of our observables with a Savitsky-Golay filter. We have tested additionally smoothing the input observables as well and found it has marginal impact on our forecasts. We note that in the case of $w_{p,vg}$ and $\Delta \Sigma_{vm}$ we remove scales below $r_p = 25.0 $ and $12.0 \, h^{-1} \, \mathrm{Mpc}$ respectively. This choice is in order to be conservative and remove scales on which our cosmological derivatives are noisy.

We describe our covariance matrix estimation methodology in the next subsection. Our forecast of the statistical error on a model parameter $\theta_i$ is $\left[ \left(F^{-1}\right)_{ii}\right]^{1/2}$ and the estimate for the covariance of two parameters $\theta_i$, $\theta_j$ is given by $\left( F^{-1} \right)_{ij}$.

\subsection{Covariance Estimation}
\label{subsec:cov}

\begin{figure*}
\centering 
\includegraphics[width=1.0\textwidth]{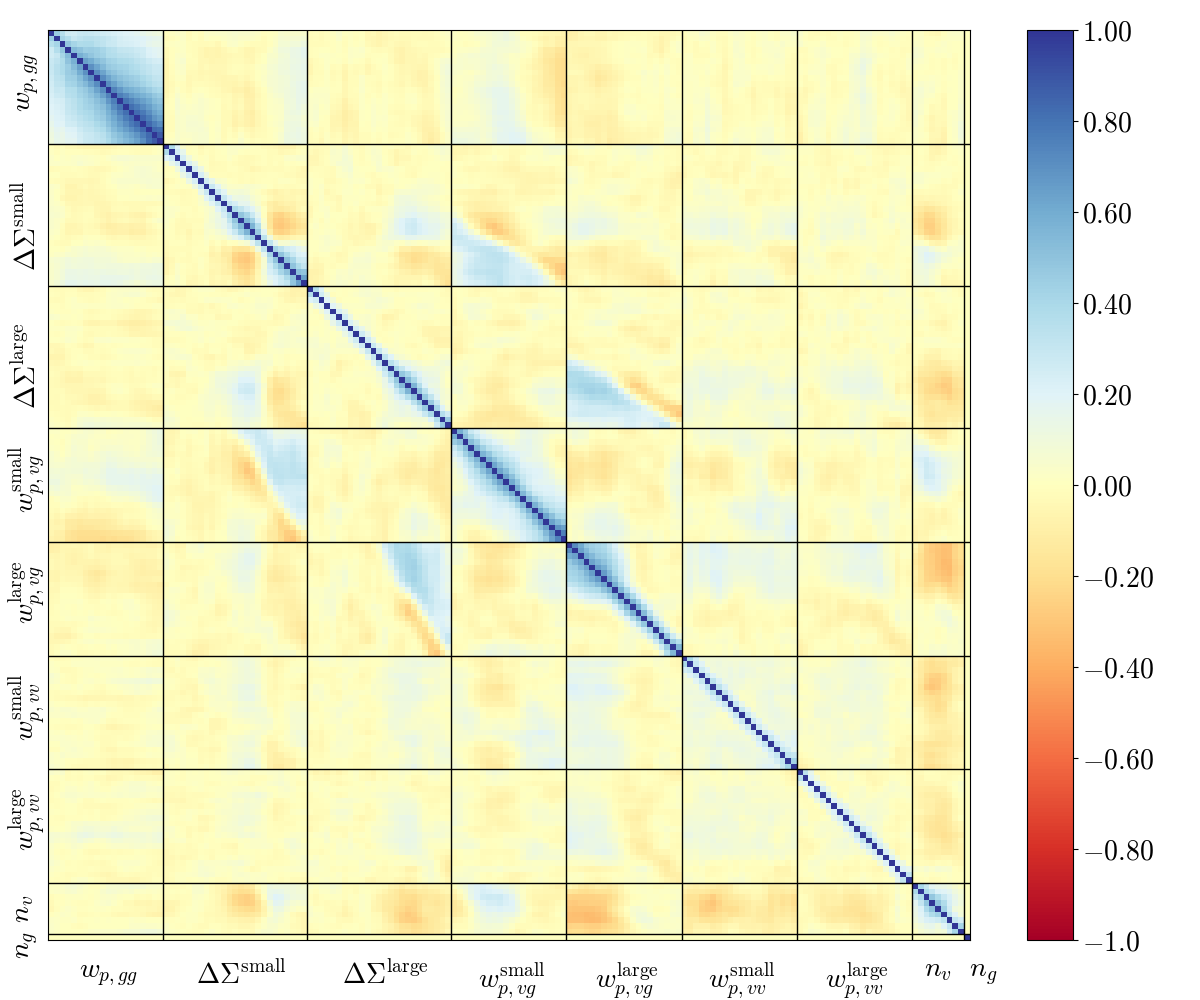}
    \caption{Numerically computed correlation matrix of all of our observables with boxcar smoothing. Correlation blocks from left to right correspond to the projected galaxy autocorrelation function, weak lensing of small voids ($\Reff \in [20, 60)$), weak lensing of large voids ($\Reff \in [60,100)$), projected small void-galaxy cross-correlation function, projected large void-galaxy cross-correlation function, projected small void autocorrelation function, projected large void autocorrelation function, void size function, and galaxy number density. The resulting covariance matrix is mostly diagonal (with the exception of $w_{p,gg}$) with minimal cross-observable covariance.
    } 
\label{fig:corr}
\end{figure*}

Covariance matrices for cosmological observables have been computed using analytic and numerical methods. Numerical methods typically rely on many realizations of a simulated dataset constituting a large volume relative to the survey they are designed to mimic and are therefore limited by computational cost. Analytic methods are noiseless and typically more computationally efficient. 

Voids represent a particularly challenging case of covariance estimation as their large range of sizes demands particularly large volumes for numerical estimation, and the ambiguities and difficulties in measuring their bias frustrate analytic estimation. To illustrate we consider the standard Gaussian covariance \citep[e.g.][]{Marian_Smith_Angulo_2015, Krause_Eifler_2017} which requires various input power spectra. For example to compute the covariance for $w_{p,cg}$ we must estimate $P_{gg}$, $P_{vv}$ and $P_{vg}$ which require measurements of the galaxy and void bias $b_g$ and $b_v$ as well as the cross-correlation coefficient between voids and galaxies $r_{vg}$. These are both extremely challenging to measure,
\beq
b_v = \sqrt{\frac{\xi_{vv}}{\xi_{mm}}},
\eeq
\beq
r_{vg} = \frac{1}{b_v b_g} \frac{\xi_{vg}}{\xi_{mm}} = \frac{r_{vm}}{b_g} \frac{\xi_{vg}}{\xi_{vm}},
\eeq
as they ultimately require precise measurements of $\xi_{vv}$ which is quite noisy (see Appendix \ref{app:wpvv}). Typically one would measure cross-correlation functions with the matter or a high density tracer, and by assuming that on large scales $r\sim1$ estimate the bias. However it is unclear whether this assumption holds in the case of voids as they are defined by their anti-correlation with some tracer, and the only way to numerically check is to precisely measure $\xi_{vv}$. Therefore we instead choose to numerically measure our covariance but note that this challenge remains in estimating and validating the covariance matrices of void observables.

To numerically estimate a measurement covariance matrix for our void observables we use subvolumes of the 25 AbacusSummit realizations of our fiducial cosmology. Each realization is divided into 16 equal volume regions by tiling a face of the box. The corresponding 400 subvolumes are rectangular prisms measuring $0.5 \times 0.5 \times 2.0 \, h^{-3} \, \mathrm{Gpc}^3$ in volume, where the major axis is taken to be the line of sight. In each subvolume we compute the observables and include pairs that cross the subvolume boundaries weighted by 0.5 \citep{Friedrich_et_al_2016}. 

To compute the numerical covariance for datavector $\mathcal{O}$ we use a bootstrap method \citep[e.g.][]{Norberg_et_al_2009}. We sample $N_\mathrm{samp.} = 400$ times with replacement from our $N_\mathrm{sub} = 400$ subvolumes and average the result to define a bootstrap sample,
\beq
\hat{\mathcal{O}}_i = \frac{1}{N_\mathrm{sub}} \sum^{N_\mathrm{sub}}_{j=1} \mathcal{O}_{R^i_j},
\eeq
where $R^i_j$ is the $j$-th element of the $i$-th random sampling of $\left[ 1, 2, ... , N_\mathrm{sub}\right]$ with replacement. This process is repeated $N_\mathrm{samp} = 400$ times to generate 400 bootstrap observables. The observable covariance is then calculated for a survey of volume $V_s$ by,
\begin{widetext}
\beq
\mathrm{cov}_\mathcal{O}(r_i, r_j) = \frac{V_\mathrm{sub}}{V_s} \sum^{N_\mathrm{samp.}}_{k=1} \left( \hat{\mathcal{O}}_k(r_i) - \langle \hat{\mathcal{O}}(r_i) \rangle \right) \left( \hat{\mathcal{O}}(r_j) - \langle \hat{\mathcal{O}}(r_j) \rangle \right),
\eeq
\end{widetext}
where $V_\mathrm{sub}$ is the volume of the individual subvolumes used to measure the covariance. The resulting covariance will be noisy, and therefore can artificially bias forecasts of parameter constraints to be too optimistic. To mitigate this possibility we additionally boxcar smooth the correlation matrix,
\beq
\mathrm{corr}_\mathcal{O}(r_i, r_j) = \frac{\mathrm{cov}_\mathcal{O}(r_i, r_j)}{\sqrt{\mathrm{cov}_\mathcal{O}(r_i, r_i)\mathrm{cov}_\mathcal{O}(r_j, r_j)}}
\eeq
such that each non-diagonal element is the average of its neighbors (excluding neighbors across the diagonal). The resulting covariance matrix is used in our forecasts.

Figure \ref{fig:corr} shows the correlation matrix of all the observables we consider including all cross-observable correlations. From left to right the observables considered are: the projected galaxy clustering, the lensing of small and large voids, the projected cross-correlation of small and large voids with the galaxies, the projected auto-correlation small and large voids, the void size function and the galaxy number density. In general we observe minimal cross-correlations between probes. We note that in Figure \ref{fig:corr} we only show the sample variance contribution to our lensing covariance, shape-noise will relatively suppress off-diagonal features. 

\subsection{Fiducial Scenario}
\label{subsec:fiducial}

\begin{figure*}
\centering 
\includegraphics[width=1.0\textwidth]{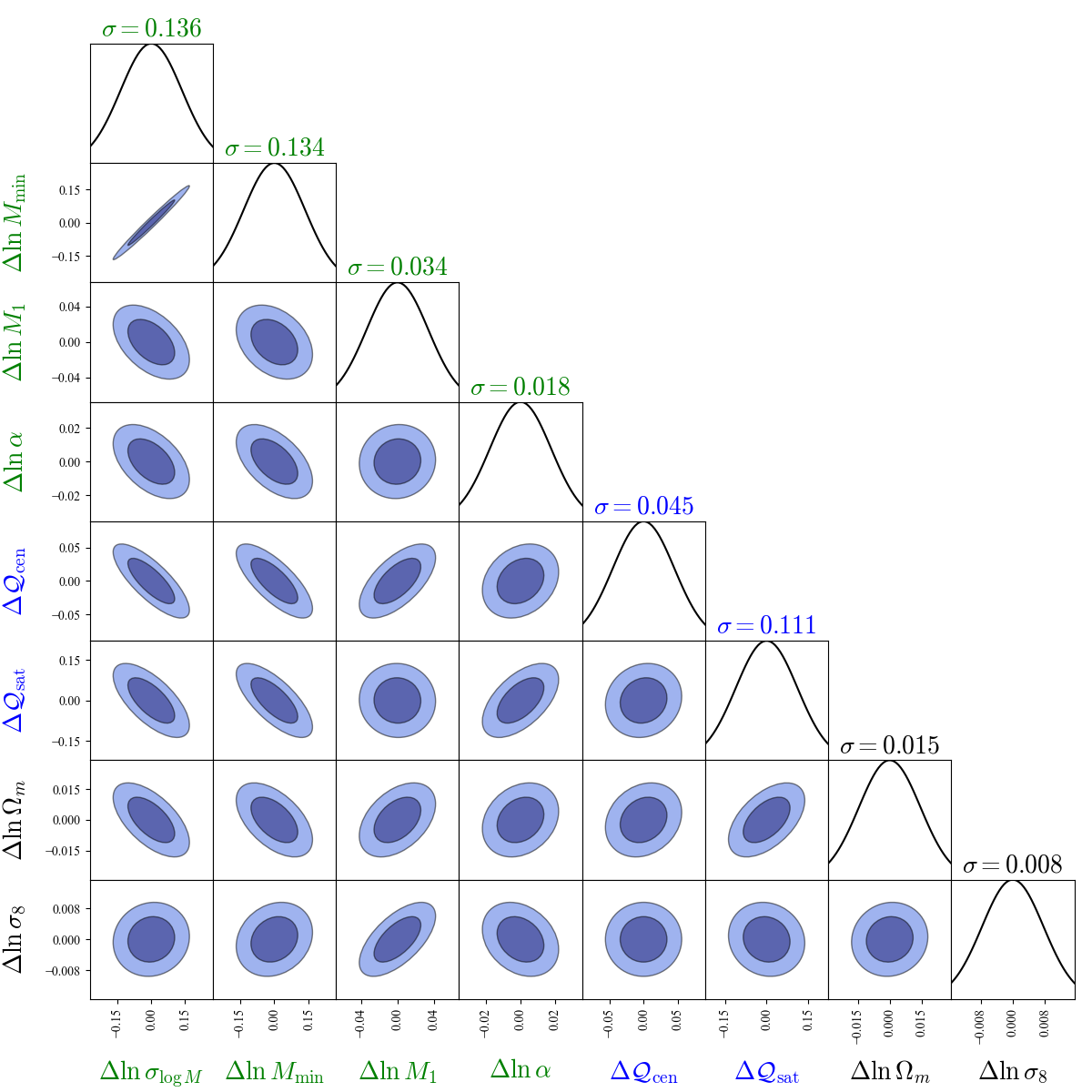}
    \caption{Forecast parameter constraints (68\% and 95\%) for our fiducial scenario, assuming DESI-Y5 survey parameters for LRG galaxies between $z = 0.4{-}0.6$ combining the void size function, projected galaxy correlation function, and projected void-galaxy cross-correlation function. We marginalize over standard HOD (green), galaxy assembly bias (blue) and cosmological (black) parameters. Fully marginalized errors on each parameter are listed above each PDF panel.} 
\label{fig:fid-corner}
\end{figure*}

We forecast parameter constraints for our fiducial scenario combining the void size function $n_v(\Reff)$, projected void-galaxy cross-correlation function $w_{p,vg}$, the projected galaxy auto-correlation function $w_{p,gg}$, and the galaxy number density $n_g$ measured in a DESI-like survey of $14000 \, \mathrm{deg}^2$ in the redshift range $0.4 < z < 0.6$\footnote{Note that full DESI data covers a larger range in redshift. The methodology we present is in principle applicable to this larger redshift range.}, with the covariance matrix described in section \ref{subsec:cov} and derivatives described in section \ref{subsec:fisher}. For our void observables we scale down our fiducial survey area by 10\% to $12600 \, \mathrm{deg}^2$ in order to account for the impact of the survey mask on void identification. In the case of DESI specifically, because the final survey mask consists of two large contiguous areas this is likely to be a conservative choice. We include cross-observable covariances (off block-diagonal elements in Figure \ref{fig:corr}) in our forecasts, though we observe that they are generally weak. Note that forecast parameters are in terms of the natural logarithm of the parameter of interest, except for the galaxy assembly bias parameters $\Qcen$ and $\Qsat$ which can plausibly achieve non-positive values. Figure \ref{fig:fid-corner} and the first row of Table \ref{table:uncertainties} present results for this fiducial case. In this case the void size function includes information from $10.0 \, h^{-1} \, \mathrm{Mpc} < \Reff < 100.0 \, h^{-1} \mathrm{Mpc}$, the projected void-galaxy cross-correlation function includes scales from $25.0 < r_p < 150.0 \, h^{-1} \, \mathrm{Mpc}$, and the projected galaxy auto-correlation function includes scales from $0.3 < r_p < 30.0 \, h^{-1} \, \mathrm{Mpc}$.

The best constrained combination of our cosmological parameters $\sigma_8$ and $\Omega_m$ is $\sigma_8 \Omega_m^{-0.038}$, with a $1\sigma$ uncertainty of 0.8\% after marginalizing over the halo-galaxy connection. Individual marginalized constraints on $\sigma_8$ and $\Omega_m$ are 0.8\% and 1.5\%. We note that our chosen combination of observables breaks the degeneracy between $\sigma_8$ and $\Omega_m$. As Table \ref{table:uncertainties} shows the $\Omega_m{-}\sigma_8$ degeneracy directions for the void size function and galaxy clustering are nearly perpendicular. This is driven by the unique response of the void size function to $\Omega_m$ as a function of void size; as shown in Figure \ref{fig:VSF_vary}: $\Omega_m$ impacts small and large voids in opposite ways. Between cosmological and HOD parameters we see a significant degeneracy between $\log M_1$ and $\sigma_8$ as well as between $\Omega_m$ and $\siglogM$, $\log \Mmin$, and $\Qsat$. In the case of $\Qsat$ and $\sigma_8$ this degeneracy appears to be introduced by $w_{p,gg}$ and is due to the fact that $\Qsat$ primarily impacts large scales of the galaxy clustering, while $\sigma_8$ predicts large scales. Therefore the combination of the two parameters can exhibit three-way degeneracies with other parameters that lack such scale dependence. The degeneracies exhibited between $\Omega_m$ and HOD parameters appear to be driven by both the void size function and galaxy clustering.

We find that our fiducial datavector produces tight constraints on our HOD parameters. The parameters $\siglogM$ and $\log \Mmin$ exhibit a significant degeneracy, but are both fairly well constrained at the 13.6\% and 13.4\% level respectively. The parameter $\siglogM$ in particular is often poorly constrained by galaxy clustering alone. We also see encouragingly tight forecast constraints on our galaxy assembly bias parameters $\Qcen$ and $\Qsat$. These results represent the first joint forecast constraints on void summary statistics and galaxy clustering that marginalizes over the galaxy-halo connection.

\begin{table*}
   \centering
   \caption{Parameter forecast uncertainties on HOD and cosmological parameters from combinations of the void size function $n_v$, projected void-galaxy cross-correlation function $w_{p,vg}$, and projected galaxy auto-correlation function $w_{p,gg}$. Entries in the first through third columns refer to the elements of the datavector that are retained. Large scales refers to $r_p > 3.0 \, h^{-1} \, \mathrm{Mpc}$, while Large/Small voids refer to voids with $\Reff\in[20,60)$/$\Reff\in[60, 100)$. All cases include the galaxy number density $n_g$ as part of the datavector.} 
   \resizebox{\textwidth}{!}{\begin{tabular}{lllccccccccccc}
\hline
$n_v$ & $w_{p,vg}$ & $w_{p,gg}$ & $\Delta \ln \sigma_{\log M}$ & $\Delta \ln M_{\mathrm{min}}$ & $\Delta \ln M_1$ & $\Delta \ln \alpha$ & $\Delta \mathcal{Q}_\mathrm{cen}$ & $\Delta \mathcal{Q}_\mathrm{sat}$ & $\Delta \ln \Omega_m$ & $\Delta \ln \sigma_8$ & $\Delta \ln \sigma_8 \Omega_m^p \left[p\right]$\\
\hline
$\checkmark$ & $\checkmark$ & $\checkmark$ & 0.136 & 0.134 & 0.034 & 0.018 & 0.045 & 0.111 & 0.015 & 0.008 &0.008 $\left[ -0.038\right]$\\
\hline
$\times$ & $\times$ & $\checkmark$ & 0.665 & 1.256 & 0.729 & 0.035 & 0.114 & 0.570 & 0.048 & 0.073 &0.060 $\left[ -0.897\right]$\\
$\checkmark$ & $\times$ & $\times$ & 3.375 & 4.016 & 11.793 & 7.244 & 0.824 & 4.787 & 0.059 & 0.155 &0.148 $\left[ 0.788\right]$\\
$\times$ & $\checkmark$ & $\times$ & 0.757 & 0.865 & 3.915 & 0.535 & 0.339 & 0.833 & 0.069 & 0.020 &0.016 $\left[ -0.177\right]$\\
\hline
$\checkmark$ & $\times$ & $\checkmark$ & 0.207 & 0.206 & 0.037 & 0.022 & 0.059 & 0.144 & 0.019 & 0.011 &0.011 $\left[ 0.137\right]$\\
$\times$ & $\checkmark$ & $\checkmark$ & 0.237 & 0.276 & 0.079 & 0.021 & 0.077 & 0.217 & 0.020 & 0.014 &0.014 $\left[ -0.119\right]$\\
$\checkmark$ & $\checkmark$ & $\times$ & 0.172 & 0.199 & 0.655 & 0.386 & 0.132 & 0.477 & 0.028 & 0.011 &0.010 $\left[ -0.151\right]$\\
\hline
$\times$ & $\times$ & Large Scales & 6.580 & 11.483 & 5.060 & 5.895 & 1.782 & 5.568 & 0.190 & 0.278 &0.090 $\left[ -1.384\right]$\\
$\checkmark$ & $\times$ & Large Scales & 0.639 & 0.563 & 0.540 & 0.794 & 0.362 & 0.823 & 0.026 & 0.019 &0.019 $\left[ -0.027\right]$\\
$\times$ & $\checkmark$ & Large Scales & 0.572 & 0.669 & 0.534 & 0.332 & 0.166 & 0.291 & 0.028 & 0.015 &0.014 $\left[ -0.195\right]$\\
\hline
Large Voids & Large Voids & $\checkmark$ & 0.270 & 0.345 & 0.137 & 0.025 & 0.074 & 0.287 & 0.025 & 0.014 &0.013 $\left[ 0.152\right]$\\
Small Voids & Small Voids & $\checkmark$ & 0.154 & 0.151 & 0.038 & 0.020 & 0.065 & 0.232 & 0.019 & 0.012 &0.011 $\left[ -0.220\right]$\\
\hline
\end{tabular}}
   \label{table:uncertainties}
\end{table*}

\subsection{Relative contributions of observables}
\label{subsec:rel_forecasts}

Table \ref{table:uncertainties} includes a variety of alternative scenarios in which we omit different elements of the fiducial datavector. In all cases we include the galaxy number density $n_g$. In some cases we retain only the large scales of the galaxy clustering, referring to $r_p > 3.0 \, h^{-1} \, \mathrm{Mpc}$. There are also cases in which we retain only the large or small bin of the void size function and void-galaxy cross-correlation. The last column of Table \ref{table:uncertainties} lists the best constrained parameter combination of $\Omega_m$ and $\sigma_8$ as well as the constraint on this combination. The first line of Table \ref{table:uncertainties} refers to our fiducial scenario described above.

The second line of Table \ref{table:uncertainties} shows a forecast with only $w_{p,gg}$ down to small scales. We see that in all cases our forecast parameter constraints have degraded significantly, with the least affected being $\alpha$ which is degraded by a factor of 2. Our cosmological parameters $\Omega_m$ and $\sigma_8$ are constrained at the 4.8\% and 7.3\% levels respectively. We see that this degradation is due in part to the significant degeneracy between $\sigma_8$ and $\Omega_m$ indicated by the index on $\Omega_m$ in their best constrained combination, as well as degeneracies between central and satellite HOD parameters. 

We next consider the case of $n_v$ alone and observe even more significant degradation of parameter constraints relative to our fiducial scenario. In the case of parameters that control the satellite occupation, $M_1$, $\alpha$, and $\Qsat$ the void size function simply has no information to constrain their values. In the case of $\Mmin$ and $\siglogM$ the parameters are almost totally degenerate with each other. The void size function constrains our cosmological parameters $\Omega_m$ and $\sigma_8$ to 5.9\% and 15.5\%, with a significant degeneracy between the two. 

In the case of the $w_{p,vg}$ alone we similarly observe significant degradation of HOD constraints relative to our fiducial scenario, though in contrast with $n_v$ we do see some constraining power on satellite parameters since $w_{p,vg}$ is sensitive to the galaxy bias. Interestingly we observe some sensitivity to cosmology with a $6.9\%$ constraint on $\Omega_m$ and a $2.0\%$ constraint on $\sigma_8$. We can compare this forecast with the results of \citet{Fraser_et_al_2024} who develop a novel HOD-based emulator for the redshift-space monopole and quadrupole of the void-galaxy cross-correlation of BOSS-CMASS galaxies and forecast constraints of $6.6\%$ and $7.0\%$ on $\Omega_m$ and $\sigma_8$ respectively. Caution should be taken when making such a comparison as we consider the {\it{projected}} void-galaxy cross-correlation in contrast to \citet{Fraser_et_al_2024}. 

In comparison to their results our forecasts assume a similar galaxy number density but a larger volume of $2.1 \, h^{-3} \,\mathrm{Gpc}$ compared to $1.1 \, h^{-1} \, \mathrm{Gpc}$. With all else equal (including datavectors) this difference would imply an improvement of ${\sim}\sqrt{2}$ in precision for our forecasts relative to theirs. However, these differences are difficult to interpret given the difference in our datavectors, their results can take advantage of information from redshift-space distortions and realistic presence of Alcock-Paczy\'nski effects \citep[AP;][]{Alcock_Paczynski_1979}. Assuming a large projection length in $w_{p,vg}$ will mitigate the impact of the AP effect, but it will still impact void-finding in realistic redshift-space catalogs. Overall we consider our results to be potentially optimistic but reasonable in comparison to the results of \citet{Fraser_et_al_2024}.

We next turn to different pairwise combinations of our datavector, beginning with the combination of the void size function $n_v$ and galaxy auto-correlation function $w_{p,gg}$. Relative to either of the individual datavectors we observe significant improvements in constraining power in both HOD and cosmological parameters. In particular we obtain $1.9\%$ and $1.1\%$ constraints on $\Omega_m$ and $\sigma_8$ respectively with little degeneracy between the two. This combination of datavectors self-calibrates the void-galaxy connection and results in significant degeneracy breaking between HOD parameters, in particular those that control satellite versus central occupation. 

Additionally the turnover in the derivative of $n_v$ we observe in Figure \ref{fig:VSF_vary} helps to break the significant $\Omega_m{-}\sigma_8$ degeneracy we observe when using galaxy clustering alone. This degeneracy breaking is crucial to the improvement we observe in constraining power, simple propagation of individual uncertainties on $\sigma_8$ from $w_{p,gg}$ and $n_v$ yields an expected $6.6\%$ constraint. We also note that our constraints on HOD parameters suggest this datavector is promising as a way to constrain models of galaxy-formation with large-scale structure information. In particular we observe significant improvements in $\siglogM$ and our central assembly bias parameters $\Qcen$. 

Our constraints for the combination of $w_{p,gg}$ and $w_{p,vg}$ are similar or slightly weaker for HOD and cosmological parameters. We obtain a $2.0\%$ constraint on $\Omega_m$, which is a significant improvement relative to either observable alone as well as the $3.9\%$ constraint one expects from simple uncertainty propagation. On the other hand we obtain a constraint of $1.4\%$ on $\sigma_8$ which is only a marginal improvement on the $2.0\%$ constraint from $w_{p,vg}$ alone. Finally we include constraints from the combination of $n_v$ and $w_{p,vg}$ for the sake of completeness. These constraints are similar but weaker compared to either of the former pairwise combinations. Additionally, because we adopt a HOD framework to predict both observables it would always make sense to include at least the large scales of galaxy clustering in any scenario that utilizes either.

We next examine the impact of removing the small-scales ($r_p < 3.0 \, h^{-1} \, \mathrm{Mpc}$) of $w_{p,gg}$ from our forecasts. In the case of $w_{p,gg}$ alone this leads to significant degradation of constraints. This highlights the importance of achieving high number density of tracer galaxies to beat down shot-noise at small scales, which also has the effect of enabling small voids to be identified in order to probe void substructure. 

Individual HOD parameters are practically unconstrained due to large scale degeneracies and this in turns leads to significant degradation of cosmological parameters relative to the all-scales case, roughly a factor of $4$ in the case of both $\Omega_m$ and $\sigma_8$. When combining the large-scales of $w_{p,gg}$ with either $n_v$ or $w_{p,vg}$ we observe much less degradation in overall constraining power. Critically, both $n_v$ and $w_{p,vg}$ help to break degeneracies between HOD parameters when combined with $w_{p,gg}$. Including small-scales and large-scale galaxy clustering and $n_v$ improves constraints on $\Omega_m$ and $\sigma_8$ by factors of roughly $\sim1.4$ and $\sim1.7$ respectively. In the case of $w_{p,vg}$ rather than $n_v$ this improvement is by a factor of $\sim1$ and $\sim1.4$ respectively.

\begin{figure}
\centering 
\includegraphics[width=0.45\textwidth]{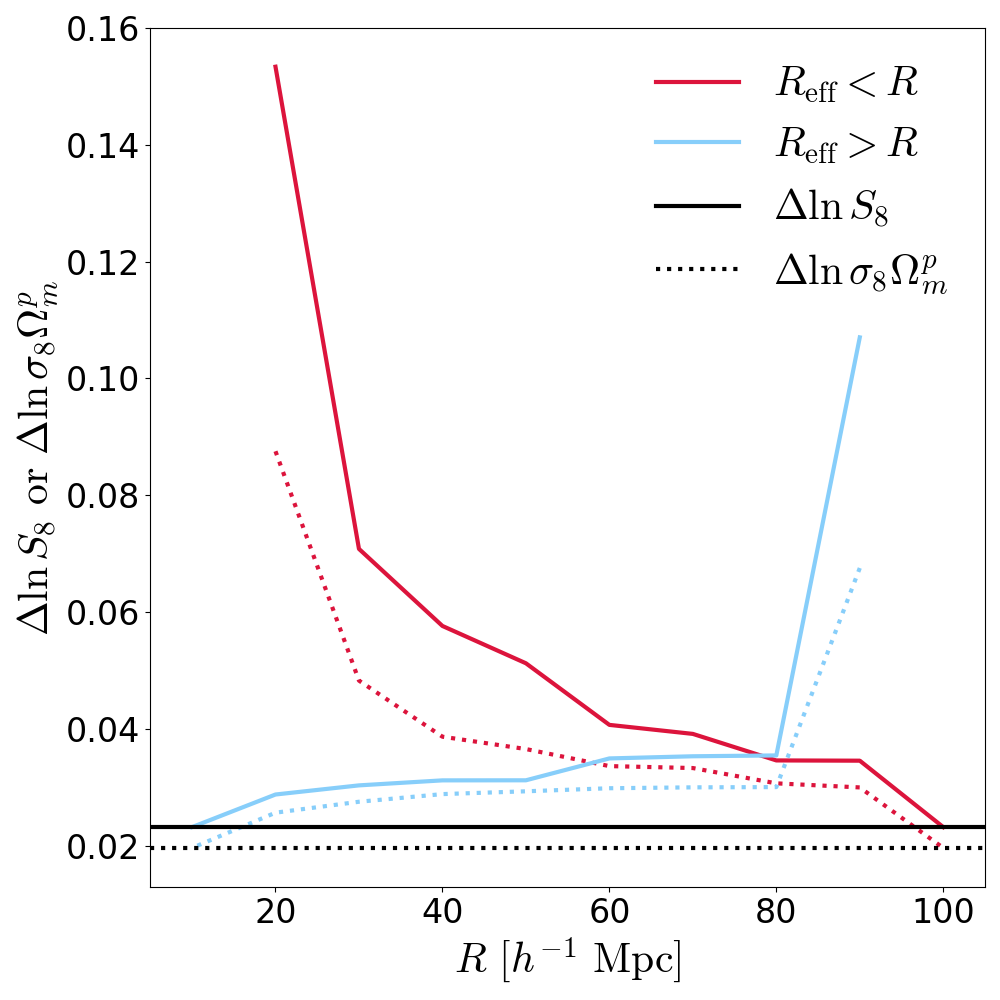}
    \caption{Forecast constraints from the combination of the void size function and galaxy clustering on either $S_8 = \sigma_8 \Omega_m^{0.5}$ (solid lines) or the best constrained combination of $\sigma_8$ and $\Omega_m$ (dashed lines) as a function of either minimum (blue) or maximum (red) void size. Horizontal lines indicate the constraint from all bins of the void size function and galaxy clustering. Constraints are fully marginalized over HOD parameters.} 
\label{fig:vary-void-size}
\end{figure}

Finally we examine the impact of void sizes on our constraints by removing either large ($\Reff \in [60, 100) \, h^{-1} \, \mathrm{Mpc}$) or small ($\Reff \in [20, 60) \, h^{-1} \, \mathrm{Mpc}$) voids from measurements of $n_v$ and $w_{p,vg}$ relative to our fiducial scenario. When removing the small voids we obtain constraints of $2.5\%$ and $1.4\%$ on $\Omega_m$ and $\sigma_8$ relative to analogous constraints of $1.5\%$ and $0.8\%$ for our fiducial scenario. When instead the large voids are removed these constraints are $1.9\%$ and $1.2\%$. These constraints are non-negligibly degraded relative to our fiducial scenario but are still interestingly competitive. We also note that the large and small voids are similarly informative when considering the combination of $n_v$ and $w_{p,vg}$. Figure \ref{fig:vary-void-size} shows a more detailed comparison of the constraints on either $S_8 \equiv \sigma_8 \Omega_m^{0.5}$ (solid) or the best constrained combination of $\Omega_m$ and $\sigma_8$ (dashed) obtained from small versus large voids when using the combination of $n_v$ and $w_{p,gg}$ only. Red curves show constraints obtained when including increasingly large voids, while blue curves show constraints obtained when including increasingly small voids.

We see that information from smaller versus larger voids is equalized at roughly $\Reff = 80.0 \, h^{-1} \, \mathrm{Mpc}$ for the combination of $n_v$ and $w_{p,gg}$. We note that this is qualitatively different than our results for $n_v$, $w_{p,gg}$ and $w_{p,vg}$ (see Table \ref{table:uncertainties}) where the small voids provide more information, likely due to the superior precision of $w_{p,vg}$ for small voids. We also observe in Figure \ref{fig:vary-void-size} that the cosmological information has not fully saturated for either the small or large voids.

\subsection{Forecasts for void-lensing}
\label{sec:lens}

The weak lensing signal of stacked voids has the potential to be a powerful probe of cosmology \cite[e.g.][]{Davies_et_al_2021, Bonici_et_al_2023}. In order to compute realistic forecasts of void-lensing we must include a shape noise-contribution to our covariance. To do this we add an additional diagonal covariance component,
\beq
\sigma^2_{\Delta \Sigma} = \Sigma_\mathrm{crit}^2 \frac{\sigma_\gamma^2}{N_\mathrm{src}},
\eeq
where $N_\mathrm{src}$ is the expected number of sources in the relevant radial bin, $\sigma_\gamma$ is the shape noise per galaxy, and the critical surface density
\beq
\Sigma_\mathrm{crit} = \frac{c^2}{4 \pi G} \frac{D_c(z_s)}{D_c(z_l) (D_c(z_l) - D_c(z_s) ) }
\eeq
where $D_c$ is the comoving distance, and $z_l$ and $z_s$ are the effective lens and source redshifts. We model source distributions for Stage-III- and Stage-IV-like weak lensing surveys. In either case we consider a source distribution of the form
\beq
f(z_s) \propto z^m \mathrm{exp}\left[ -( z / z_*)^\beta \right],
\eeq
motivated by \citet{Rozo_2011}. For Stage-III forecasts we assume $z_* = 0.5$, $m=2$, and $\beta = 1.4$ \citep{Rozo_2011} with a total source density of $\Sigma_s = 10.0 \; \mathrm{arcmin}^{-2}$. For Stage-IV forecasts we assume $z_* = 0.28$, $m = 2$, and $\beta = 0.90$ \citep{LSST_DESC_SRD} with a total source density of $\Sigma_s = 27.0 \; \mathrm{arcmin}^{-2}$. We use these model source distributions to compute the average source redshift and effective source density for sources $z_s > 0.6$ and in both cases we assume a shape noise per galaxy of $\sigma_\gamma = 0.25$ \citep[e.g.][]{Xu_J_KL_et_al_2023}.

We also consider forecast scenarios for Stage-IV kinematic lensing (KL) surveys. Kinematic lensing refers to the technique first proposed by \citet{Huff_et_al_2013} whereby resolved spectroscopic observations of source galaxies and the Tully-Fisher relation \citep{Tully_Fisher_1977} are used to infer intrinsic galaxy shapes. By thereby breaking the shape-shear degeneracy KL is expected to achieve shape noises per galaxy of $\sigma_\gamma = 0.02{-}0.04$. Additionally KL is robust to intrinsic alignments \citep[e.g.][]{RS_Pranjal_et_al_2023, Huang_et_al_2024} and bypasses the need for modeling source photometric redshift distributions. For a Stage-IV-like KL survey we assume that negligible sources are below $z = 0.6$ and therefore we will adopt our assumed total source densities as effective source densities, we additionally assume an average source redshift of $z_s = 1.0$ \citep{Xu_J_KL_et_al_2023}. 

\begin{figure}
\centering 
\includegraphics[width=0.45\textwidth]{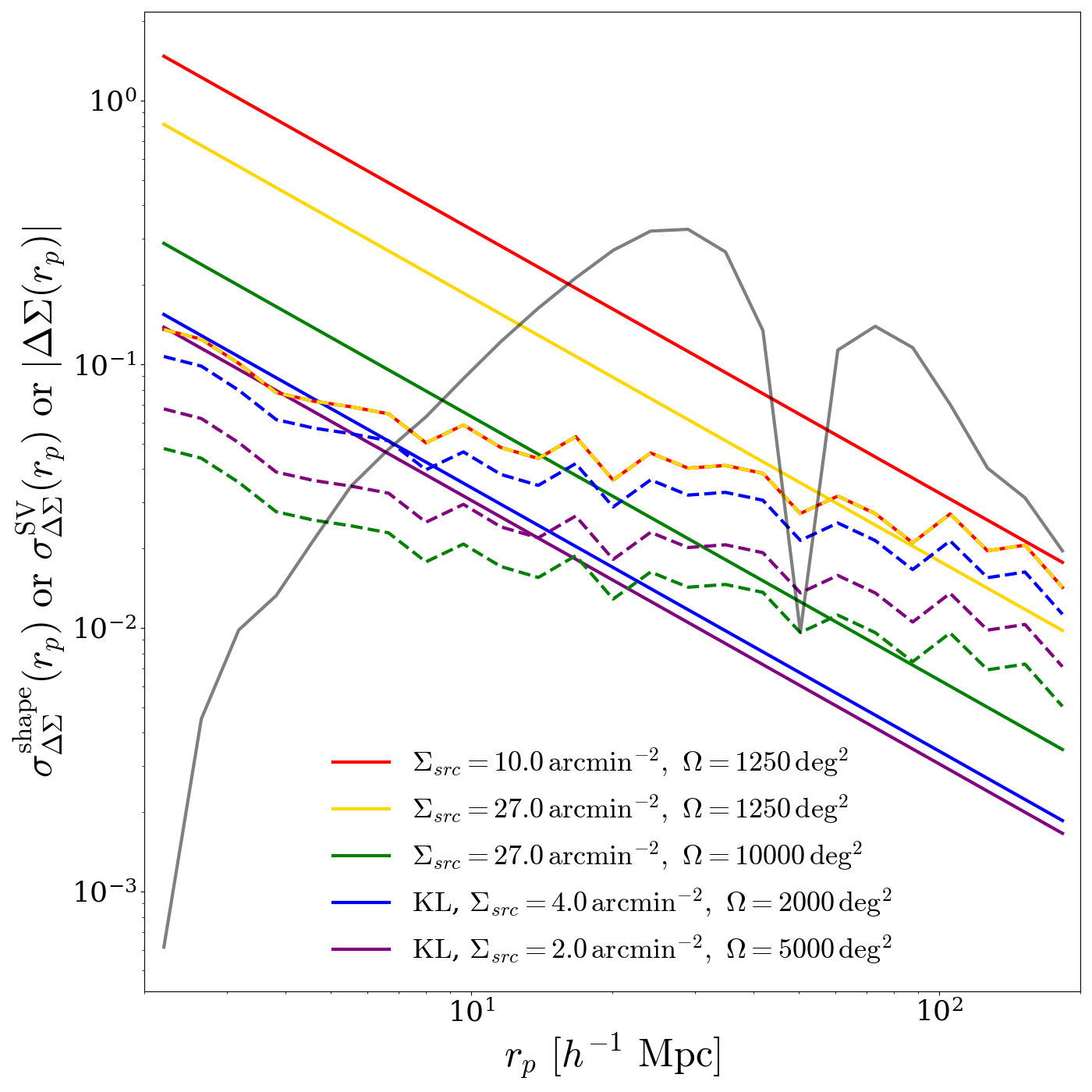}
    \caption{Comparison of the absolute value of our fiducial void lensing signal (grey line), shape noise for different survey assumptions (colored lines) and sample variance for different survey assumptions (colored dashed lines).} 
\label{fig:shape}
\end{figure}

\begin{table*}
   \centering
   \caption{Parameter forecast uncertainties on cosmological parameters from the void size function and void lensing for different lensing survey assumptions. All constraints are fully marginalized over HOD parameters.}
   \begin{tabular}{lccccccc}
      \hline
        Survey& $\Omega$ $\left[ \mathrm{deg}^2 \right]$ & $\sigma_\gamma$ & $\Sigma_\mathrm{src}$ $\left[ \mathrm{arcmin}^{-2} \right]$ & $\langle z_s \rangle$ & $\Delta \ln \Omega_m$ & $\Delta \ln \sigma_8$ & $\Delta \ln \sigma_8 \Omega_m^p \left[p\right]$\\
 	\hline
    VSF Only & - & - & - & - & 0.059 & 0.155 & 0.148 $\left[ 0.788\right]$ \\
    \hline
    Stage III  & 1250  & 0.25  & 10.0 & 1.02 & 0.052 & 0.035 &  0.034 $\left[0.082\right]$ \\
    Stage IV & 1250  & 0.25  & 27.0 & 1.15 & 0.048 & 0.026 &  0.026 $\left[0.067\right]$ \\
    Stage IV (Optimistic overlap) & 10000 & 0.25  & 27.0 & 1.15 & 0.037 & 0.015 &  0.015 $\left[0.010\right]$ \\
    Stage IV KL & 2000  & 0.025 & 4.0 & 1.00 & 0.040 & 0.017 &  0.017 $\left[0.037\right]$ \\
    Stage IV Wide KL & 5000 & 0.025 & 2.0 & 1.00 & 0.033 & 0.012 & 0.012 $\left[0.012\right]$ \\
    \hline        
    Stage IV Wide KL Only & - & - & - & - & 0.075 & 0.026 & 0.024 $\left[-0.114\right]$ \\
    \hline
   \end{tabular}
   \label{table:lens-scenarios}
\end{table*}

In Figure \ref{fig:shape} we consider five possible survey scenarios and compare the expected shape noise (colored solid lines) for each to the diagonal sample variance  contribution to the covariance (colored dashed lines) and signal (gray line) for the lensing of our small void bin ($\Reff \in [20, 60)$). The red and gold lines correspond to Stage-III and Stage-IV surveys respectively with an assumed overlap of $1250 \, \mathrm{deg}^2$ with DESI. The green line shows the impact of increasing the overlap of the Stage-IV survey to an optimistic $10,000 \, \mathrm{deg}^2$. Finally the blue and purple line shows the expected shape noise for possible Stage-IV KL surveys. The blue line corresponds to a survey with $2000 \, \mathrm{deg}^2$ of overlap while the purple corresponds with $5000 \, \mathrm{deg}^2$ of overlap. These scenarios are roughly motivated by the planned Nancy Grace Roman High Latitude survey \citep[e.g.][]{Eifler_et_al_2021_multiprobe} and a possible wide-field survey from Roman data \citep{Eifler_et_al_2021_LSST}. Such catalogs \citep{Xu_J_KL_et_al_2023} are expected to have a source density of roughly $4.0$ and $2.0 \, \mathrm{arcmin}^{-2}$ respectively and shape noise per galaxy of $\sigma_\gamma = 0.025$. 

In Table \ref{table:lens-scenarios} we forecast cosmological constraints from the combination of the void size function and void weak lensing for different survey scenarios. Table \ref{table:lens-scenarios} only shows constraints on cosmological parameters $\Omega_m$, $\sigma_8$ and their best-constrained combination $\sigma_8 \Omega_m^p$, though all constraints are fully marginalized over HOD parameters. Overall we see that the void size function and void lensing are highly complementary and that the combination produces significant improvements in cosmological constraints relative to the void size function alone. 

For our Stage-III forecast scenario we see that the addition of void lensing to the size function modestly improves constraining power on $\Omega_m$ from $5.9\%$ to $5.2\%$ while constraints on $\sigma_8$ are improved by a factor of ${\sim}4.5$ from $15.5\%$ to $3.5\%$. With deeper imaging from Stage-IV lensing surveys at fixed overlap area these constraints improve to $4.8\%$ and $2.6\%$ respectively. Finally if the overlap between our void catalog and Stage-IV lensing were to improve from $1250\, \mathrm{deg}^2$ to $10000\, \mathrm{deg}^2$ we forecast constraints of $3.7\%$ and $1.5\%$ on $\Omega_m$ and $\sigma_8$ respectively. In this forecast scenario the degeneracy between these two parameters is negligible, with their best constrained combination $\sigma_8 \Omega_m^{0.01}$ constrained at the $1.5\%$ level. These constraints are encouragingly competitive and, additionally, because they are derived from a void-only datavector they allow for valuable systematics cross-checks with other probes.

Finally we turn to forecast constraints from our KL scenarios. In contrast to ``standard'' Stage-IV lensing these scenarios assume significantly improved shape noise, at the cost of reduced total source density. This trade-off is generally favorable, with a factor of 10 reduction in shape noise per galaxy corresponding to a factor of 100 reduction in total shape noise and a factor of several reduction in survey area only corresponds to an equivalent factor of several reduction in total shape noise. Additionally since our KL scenarios are roughly motivated by the Roman Space Telescope measurements they assume relatively higher source redshifts and therefore a large fraction of the total source density is behind our lenses. We expect additional improvements relative to ``standard'' lensing in an actual data analysis due to KL surveys being able to ignore systematics associated with intrinsic alignments and photometric redshift estimation. 

For our Stage-IV KL survey scenario we forecast $4.0\%$ and $1.7\%$ constraints on $\Omega_m$ and $\sigma_8$. For our wide KL survey scenario these constraints improve to $3.3\%$ and $1.2\%$ respectively. This improvement can be understood in the context of Figure \ref{fig:shape}, we see that with $4.0 \, \mathrm{arcmin}^{-2}$ of source KL galaxies our void-lensing datavector is already strongly sample variance dominated where we expect significant signal. Finally we include forecast constraints from our wide KL survey scenario for void lensing alone. In this case we forecast $7.5\%$ and $2.6\%$ constraints on $\Omega_m$ and $\sigma_8$. This result demonstrates the potential of void-lensing in the context of upcoming Stage-IV surveys. Particularly interesting is the possibility to constrain modified theories of gravity \citep[e.g.][]{Joyce_et_al_2015, Kazuya_MG_review_2016} by testing the consistency of void lensing and the measurements of void outflow kinematics. Such tests are similar to investigations of cluster infall kinematics \citep[e.g.][]{Zu_et_al_2014B} but may prove to be more sensitive to modified gravity since voids are less or even not screened objects. We leave investigation of such constraints to future work. 

\section{Conclusions}
\label{sec:conc}

We have investigated potential cosmological constraints from multi-probe cosmology with the precision expected in DESI Year 5 data. With DESI data alone we examined constraints from the combination of the void size function $n_v$, projected void-galaxy cross-correlation function $w_{p,vg}$, and projected galaxy auto-correlation function $w_{p,gg}$. We have additionally examined constraints from combining the void size function with void lensing measured from overlapping Stage III and Stage IV surveys. We have computed observables using simulations from the {\sc{AbacusSummit}} suite \citep{Maksimova_Summit_et_al_2021} of cosmological N-body simulations and populating haloes with mock galaxies using a flexile HOD parameterization that includes central and satellite galaxy assembly bias. Using these observables we compute derivatives of our datavector with respect to HOD and cosmological parameters and compute Fisher matrix forecasts for a variety of survey scenarios.

For our fiducial forecast we assume a DESI-like survey of galaxies from which we can identify voids. Our forecasts are for galaxies and voids within the redshift range $z = 0.4{-}0.6$. We assume a survey area of $12600\,\mathrm{deg}^2$, a $10\%$ reduction relative to the expected $14000\,\mathrm{deg}^2$ in the final data release of DESI in order to account for the impact of survey mask on void identification.  We assume fiducial galaxy HOD parameters motivated by the galaxy-halo connection of luminous red galaxies (LRGs) and include central and galaxy assembly bias. To predict covariance matrices we rely on a combination of analytical and numerical methods.

Our fiducial forecast combining the void size function $n_v$, projected void-galaxy cross-correlation function $w_{p,vg}$, and projected galaxy auto-correlation function $w_{p,gg}$ yields constraints of $1.5\%$ and $0.8\%$ on $\Omega_m$ and $\sigma_8$ respectively. These constraints are enabled by degeneracy breaking between different elements of our datavector. Since void identification is negligibly impacted by the details of the satellite galaxy-halo connection the combination of our void observables with galaxy clustering helps to break internal central-satellite degeneracies within the galaxy halo connection. Meanwhile the unique response of our void observables to $\Omega_m$ (described in Section \ref{sec:sens}) breaks the degeneracy with $\sigma_8$. This is illustrated by our best constrained parameter combination $\sigma_8 \Omega_m^{-0.038}$ being constrained at the $0.8\%$ level. These constraints may improve in redshift-space due to the cosmological information contained within RSD and AP effects, though this improvement will be mitigated by the inclusion of nuisance parameters to model galaxy velocities. In practice carrying out such an analysis will require fast and accurate emulation of our model datavector. We leave the development of such emulators for future work but note that the literature amply demonstrates the feasibility of this task \citep[e.g.][]{Zhai_AemulusII_2019, Wibking_et_al_2019, Yuan_et_al_2022, Zhai_et_al_2023, Fraser_et_al_2024, Lehman_et_al_2025}.

\begin{figure}
\centering 
\includegraphics[width=0.45\textwidth]{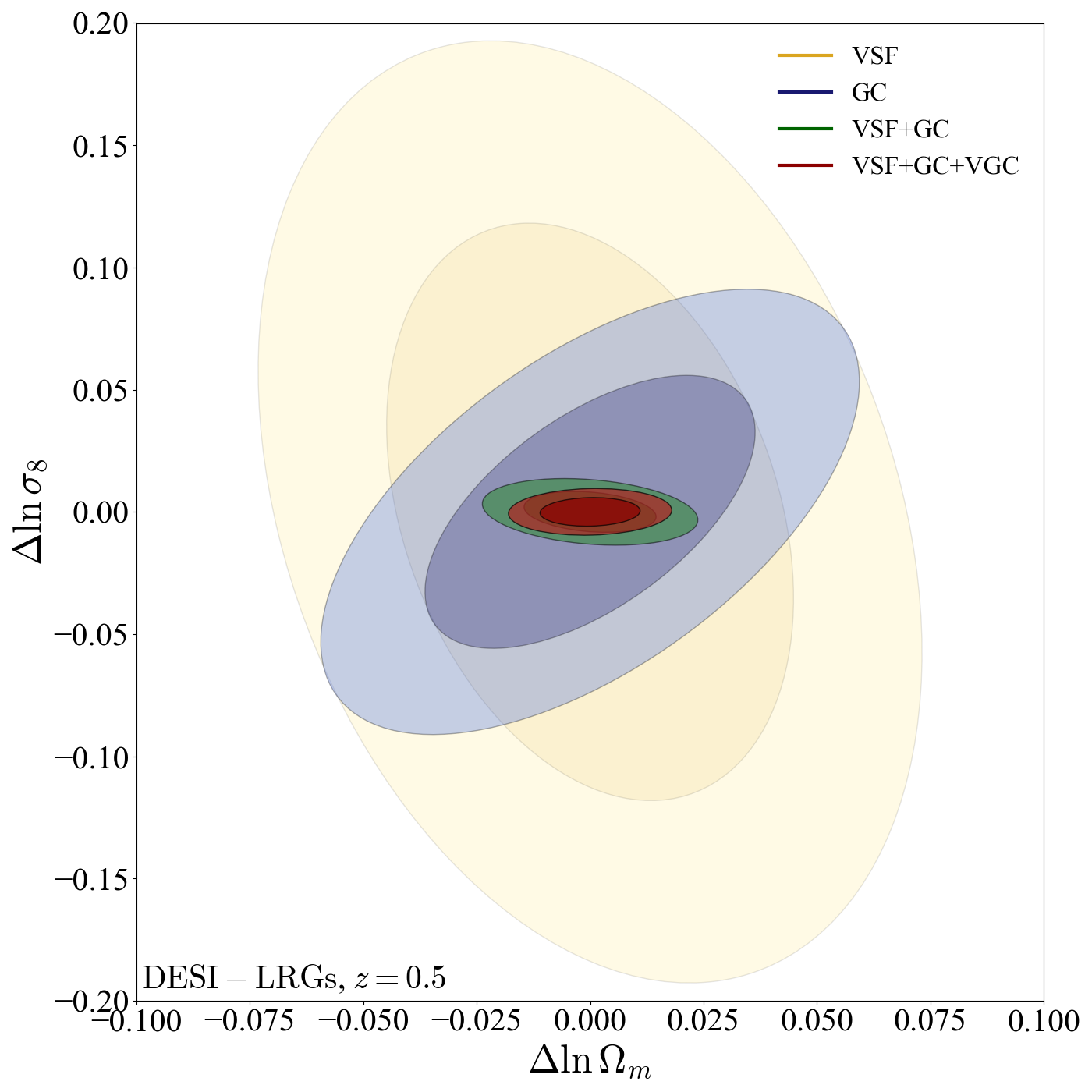}
    \caption{Forecast constraints (68\% and 95\% contours) on $\sigma_8$ and $\Omega_m$ marginalized over all HOD parameters for different observable combinations. Yellow and blue contours represent constraints from the void size function and galaxy clustering respectively. Green contours shows the result of combining the two, and the red shows the result of further including the void-galaxy cross-correlation.} 
\label{fig:multi-probe}
\end{figure}

We have also investigated the cosmological constraints we expect from various subsamples of our fiducial datavector; the main results of this investigation are summarized in Figure \ref{fig:multi-probe}. We show constraints on $\sigma_8$ and $\Omega_m$ fully marginalized over our HOD and assembly bias model. We see that constraints from either the void size function (yellow) or small and large scale galaxy clustering (blue) {\it{alone}} are relatively uninteresting in the context of Stage IV surveys. In the case of the void size function these constraints are $5.9\%$ and $15.5\%$ on $\Omega_m$ and $\sigma_8$ respectively, compared to $4.8\%$ and $7.3\%$ from galaxy clustering. Their combination however (green) significantly improves on both of these due to the degeneracy breaking described above, providing fully marginalized constraints on $\Omega_m$ and $\sigma_8$ of $1.9\%$ and $1.1\%$. Our fiducial forecast which additionally includes the void-galaxy cross-correlation (red) further improves on these constraints.

These results are of particular significance for several reasons. Galaxy clustering is a mature cosmological probe with a variety of modeling frameworks \citep[e.g.][]{Desjacques_et_al_2018, Wechsler_Tinker_rev_2018, Cabass_EFTrev_et_al_2023} available for use in constraining cosmology with existing and upcoming and spectroscopic data. Halo-model frameworks of the galaxy-halo connection such as HOD and SHAM already aim to marginalize over all that is necessary to predict the void population identified with the same galaxy sample. Therefore, within a simulation-based forward modeling context there should in principle be no additional cost in nuisance parameters to predict the void population and its various summary statistics relative to that needed to predict the galaxy clustering. Were this assumption to prove untrue it would demonstrate the ability of void statistics to provide information on the galaxy formation that is not available from galaxy clustering alone.

Additionally, although we present fiducial results that include the void-galaxy cross-correlation we emphasize that cosmological constraints from the void size function and galaxy clustering are similarly competitive. The covariance matrix for the void-galaxy cross-correlation is by far the most challenging to estimate of the three observables, and so it is encouraging that we can still obtain tight constraints from the combination of galaxy clustering, whose covariance is well studied in the literature, and the void size function, whose covariance is relatively simple.

We have also presented forecasts that combine void lensing with the void size function for a variety of Stage-III and Stage-IV lensing survey scenarios. With $1250 \, \mathrm{deg}^2$ of Stage-III imaging we forecast $5.2\%$ constraints on $\Omega_m$ and $3.5\%$ constraints on $\sigma_8$. With Stage-IV imaging these constraints improve to $4.8\%$ and $2.6\%$ respectively and further improve to $3.7\%$ and $1.5\%$ when assuming an imaging overlap of $10000 \, \mathrm{deg}^2$. We also examine constraints from Stage-IV kinematic lensing (KL), and find that with $2000 \, \mathrm{deg}^2$ of KL data we expect $4\%$ constraints on $\Omega_m$ and $1.7\%$ constraints on $\sigma_8$. Such an analysis will be feasible with Roman lensing data \citep{Xu_J_KL_et_al_2023}. These forecast constraints are interesting within the context of $\Lambda$CDM as well as various extensions including modified gravity.

Finally our results are interesting in the context of the tension in $S_8$ between late- and early-Universe probes. In the late-Universe this tension is broadly driven by the combination of measurements of weak lensing, peculiar velocities, redshift-space distortions, and cluster abundances  \citep[e.g.][]{Abdalla_et_al_2022}. Given the recent results from the KIDS collaboration \citep{Wright_KIDS_et_al_2025}, indicating a relaxation of the tension, it becomes all the more important to add independent probes such as voids to the landscape, and to do so while properly accounting for systematics (e.g. the impact of the galaxy-void connection). Our forecast constraints from voids are precise enough to further sharpen this tension relax it. Additionally, since voids are sensitive to distinct systematics, they will allow for valuable checks of consistency between late-Universe probes. 

\section*{Acknowledgements}

We thank Enrique Paillas, and the participants of the Voids@CPPM 2025 workshop for valuable discussions about this work. We thank the AbacusSummit team for making their simulations publicly available. Simulations were analyzed in part on computational resources of the Ohio Supercomputer Center \cite{OhioSupercomputerCenter1987}, with resources supported in part by the Center for Cosmology and AstroParticle Physics at the Ohio State University. We gratefully acknowledge the use of the {\sc{matplotlib}} software package \cite{Hunter_2007} and the GNU Scientific library \cite{GSL_2009}. This research has made use of the SAO/NASA Astrophysics Data System. AS was supported by the Department of Energy grant DE-SC0020215. AP acknowledges support from the European Research Council (ERC) under the European Union's Horizon programme (COSMOBEST ERC funded project, grant agreement 101078174), as well as support from the French government under the France 2030 investment plan, as part of the Initiative d'Excellence d'Aix-Marseille Université - A*MIDEX AMX-22-CEI-03.

\section*{Data Availability}
The data underlying this article will be shared on reasonable request to the corresponding author.

\bibliography{masterbib}
\bibliographystyle{apsrev}

\appendix

\section{Void autocorrelation}
\label{app:wpvv}

\begin{figure*}
\centering 
\includegraphics[width=1.0\textwidth]{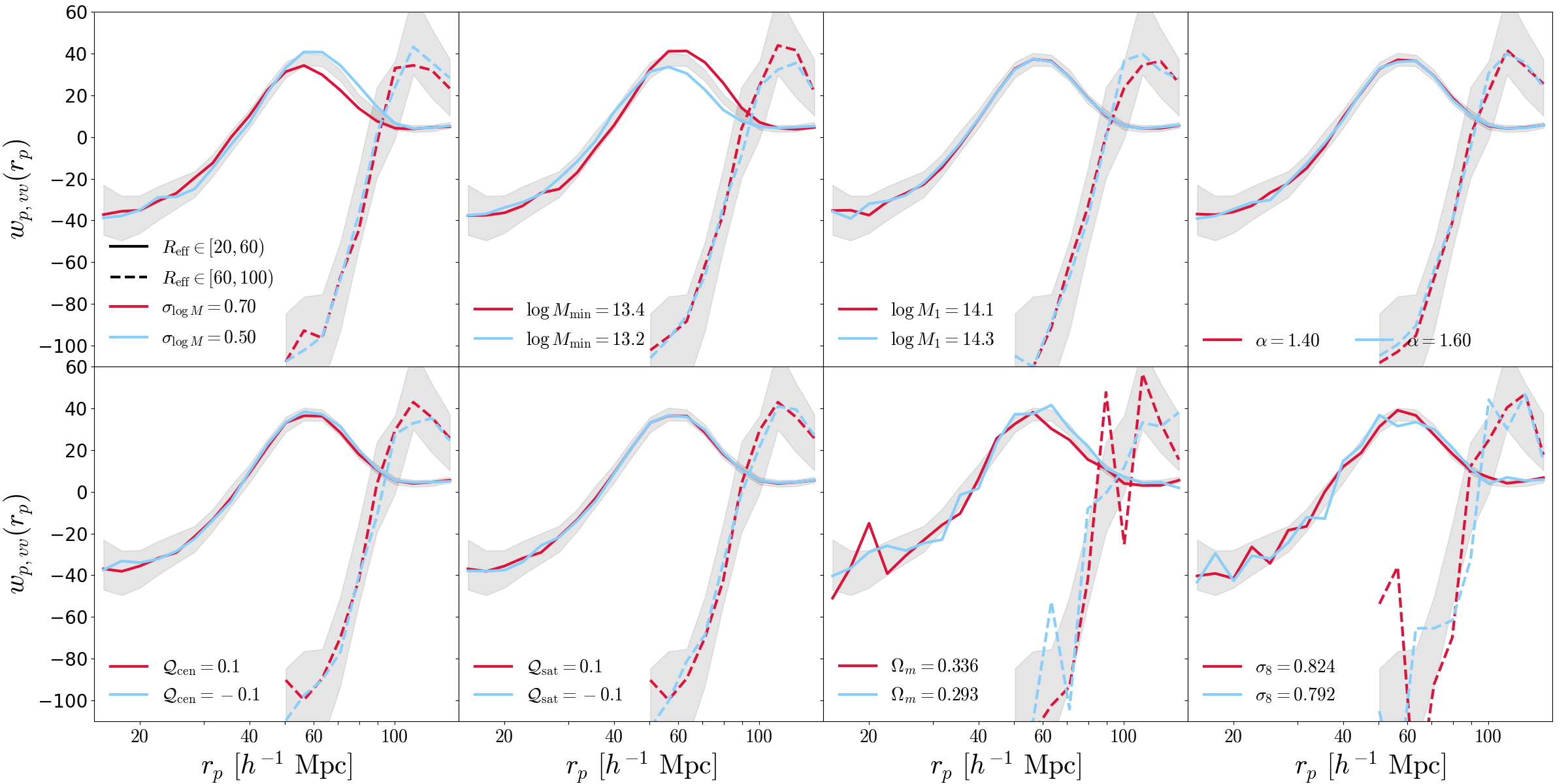}
    \caption{Sensitivity of the projected void auto-correlation function to our HOD and cosmological parameters in two bins of void size, $R_v \in [ 20, 60 \, h^{-1} \, \mathrm{Mpc})$ (solid) and $R_v \in [ 60, 100 \, h^{-1} \, \mathrm{Mpc}]$ (dashed). Each panel shows variations in an HOD or cosmological parameter around our fiducial model. Small scale bins of the large void $w_{p,vv}$ impacted by void-exclusion have been suppressed for visual clarity.
    } 
\label{fig:wpvv_vary}
\end{figure*}

Our HOD-based modeling framework is also capable of predicting the void autocorrelation function which has the potential to provide valuable cosmological constraints in upcoming surveys \citep[][]{ChanKC_et_al_2014, Kreisch_et_al_2019, Kreisch_et_al_2022}. For completeness we include sensitivity tests of the projected void autocorrelation function in Figure \ref{fig:wpvv_vary} analogous to those for the void size function, void lensing, and void-galaxy correlation function in Section \ref{sec:sens}. We note that the sensitivity of $w_{p,vv}$ to our HOD parameters is qualitatively similar to our other observables, with the exception of $\Qcen$ our central assembly bias parameter which it appears to be insensitive to. We have chosen not to include $w_{p,vv}$ in our forecasts because we do not trust our ability to accurately measure derivatives of our cosmological parameters $\Omega_m$ and $\sigma_8$ due to noise.

\end{document}